\documentclass[twoside,fleqn]{article}
\usepackage{epsfig}
\usepackage{espcrc2}

\newcommand{\AmS}{{\protect\the\textfont2
  A\kern-.1667em\lower.5ex\hbox{M}\kern-.125emS}}

\def\H{H\hskip-8.5pt/\hskip2pt}

\def\coeff#1#2{{\textstyle{#1\over #2}}}

\def\VEV#1{\left\langle #1\right\rangle}

\def\lsim{\mathrel{\mathpalette\@versim<}}

\def\gsim{\mathrel{\mathpalette\@versim>}}

\def\Tr{{\rm Tr}\,}

\title{~~~~~~~~~~~~~~~~~~~~~~~~~~~~~~~~~~~~~~~~~~~~~~~~~~~~~~~~~~~~~~~~~~~~~~~~~~~~~~~~~~~~~~~~{\small ACT-05-06,~CERN-TH/2006-144,}  \\
~~~~~~~~~~~~~~~~~~~~~~~~~~~~~~~~~~~~~~~~~~~~~~~~~~~~~~~~~~~~~~~~~~~~~~~~~~~~~~~~~~~~~~~~~~~~~{\small FTUV-06-08-01,~MIFP-06-19} \\
\vspace{0.5cm}
CPT and Quantum Mechanics Tests with Kaons }

\author{Jose Bernab\'eu\address{Departamento de F\'isica Te\'orica and IFIC,
Universidad de Valencia-CSIC,\\
E-46100 Burjassot (Valencia), Spain}, John Ellis\address{CERN,
Physics Department, Theory Division,
\\
CH-1211 Geneva 23, Switzerland}, Nick E. Mavromatos\address{King's
College London, Department
of Physics \\
Strand, London WC2R 2LS, United Kingdom},
Dimitri V. Nanopoulos\address{George P. and Cynthia W. Mitchell
Institute for Fundamental
Physics, \\ Texas A\&M
University, College Station, TX 77843, USA; \\
Astroparticle Physics Group, Houston
Advanced Research Center (HARC),
Mitchell Campus,
Woodlands, TX~77381, USA; \\
Academy of Athens, Division of Natural Sciences, 28~Panepistimiou Avenue,
Athens 10679, Greece}, and Joannis
Papavassiliou$\rm^a$}%

\begin{document}

\begin{abstract}
In this review we first discuss the theoretical motivations for
possible CPT violation and deviations from ordinary quantum-mechanical
behavior of field-theoretic systems in the context of an extended class
of quantum-gravity models. Then we proceed to a description of
precision tests of CPT symmetry using mainly neutral kaons. We
emphasize the possibly unique r\^ole of neutral meson factories in
providing specific tests of models where the quantum-mechanical
CPT operator is not well-defined, leading to modifications of
Einstein-Podolsky-Rosen particle correlators. Finally, we
present tests of CPT, T, and CP using charged
kaons, and in particular $K_{\ell 4}^\pm$ decays, which are interesting
due to the high statistics attainable in experiments.

\end{abstract}

\maketitle

\section{CPT Symmetry and Quantum Gravity: Motivations for
its Possible Violation}

Any complete theory of {\it quantum gravity} (QG) is bound to address
fundamental issues, directly related to the emergence of space-time
and its structure at energies beyond the Planck energy scale $M_P
\sim 10^{19} $ GeV. From our experience with low-energy local
quantum field theories on flat space-times, we are tempted to expect
that a theory of QG should respect most of the
fundamental symmetries that govern the standard model of electroweak
and strong interactions, specifically Lorentz symmetry and CPT
invariance, that is invariance under the combined action of Charge
Conjugation (C), Parity (P) and Time Reversal Symmetry (T).

CPT invariance is guaranteed in flat space-times by a theorem
applicable to any local quantum field theory of the type used
to describe the standard phenomenology of particle physics to date.
The \emph{ CPT theorem} can be stated as follows~\cite{cpt}:
\emph{Any quantum theory formulated on flat space-times is symmetric
under the combined action of CPT transformations, provided the
theory respects (i) Locality, (ii) Unitarity (i.e. conservation of
probability) and (iii) Lorentz invariance.}

The extension of this theorem to QG is far
from obvious. In fact, it is still a wide open and challenging issue,
linked with our (very limited at present) understanding of QG,
as well as the very nature of space-time at
(microscopic) Planckian distances $10^{-35}$~m. The important point
to notice is that the CPT theorem \emph{may not} be valid (at least
in its strong form) in highly curved (singular) space-times, such as
black holes, or more general in some QG models involving
{\it quantum space-time foam} backgrounds~\cite{wheeler}. The latter
are characterized by singular quantum fluctuations of space-time
geometry, such as black holes, {\it etc.}, with event horizons of
microscopic Planckian size. Such backgrounds result in {\it
apparent} violations of {\it unitarity} in the following sense:
there is some part of the initial information (quantum numbers of
incoming matter) which ``disappears'' inside the microscopic event
horizons, so that an observer at asymptotic infinity will have to
trace over such ``trapped'' degrees of freedom. One faces therefore
a situation where an initially pure state evolves in time and
becomes mixed. The asymptotic states are described by density
matrices, defined as
\begin{equation}
\rho _{\rm out} = {\rm Tr}_{M} |\psi ><\psi|~, \end{equation} where
the trace is over trapped (unobserved) quantum states that
disappeared inside the microscopic event horizons in the foam. Such
a non-unitary evolution makes it impossible to define a
standard quantum-mechanical scattering matrix. In ordinary local
quantum field theory, the latter connects asymptotic state vectors
in a scattering process
\begin{equation}
|{\rm out}> = S~|{\rm in}>,~S=e^{iH(t_f - t_i)}~,
\end{equation}
where $t_f - t_i$ is the duration of the scattering (assumed to be
much longer than other time scales in the problem, i.e.
${\rm lim}~t_i \to -\infty$, $t_f \to +\infty$). Instead, in
foamy situations, one can only define an operator that connects
asymptotic density matrices~\cite{hawking}:
\begin{equation}\label{dollar}
\rho_{\rm out} \equiv {\rm Tr}_{M}| {\rm out} ><{\rm out} | = \$
~\rho_{\rm in},~ \quad \$ \ne S~S^\dagger.
\end{equation}
The lack of factorization is attributed to the apparent loss of
unitarity of the effective low-energy theory, defined as the part of
the theory accessible to low-energy observers performing scattering
experiments. In such  situations particle phenomenology has to be
reformulated~\cite{ehns,poland} by viewing our low-energy world as
an open quantum system and using (\ref{dollar}). Correspondingly,
the usual Hamiltonian evolution of the wave function is replaced
by the Liouville equation for the density matrix~\cite{ehns}
\begin{equation}\label{evoleq}
\partial_t \rho = i[\rho, H] + \delta\H \rho~,
\end{equation}
where $\delta\H \rho $ is a correction of the form normally found
in open quantum-mechanical systems~\cite{lindblad}.

The \$ matrix is {\it not invertible}, and this reflects the
effective unitarity loss. It is this property that leads to a
violation of CPT invariance, since one of the requirements of CPT
theorem (unitarity) is violated. But in this particular case there
is something more than a mere violation of the symmetry. The CPT
operator itself is \emph{not well-defined}, at least from an
effective field theory point of view. This is a strong form of CPT
violation (CPTV). There is a corresponding theorem by Wald~\cite{wald}
describing the situation: \emph{In an open (effective) quantum
theory, interacting with an environment, e.g., quantum
gravitational, where} $ \$ \ne SS^\dagger $, \emph{CPT invariance is
violated, at least in its strong form}.

The proof is based on elementary quantum mechanical concepts and the
above-mentioned non-invertibility of \$, as well as the relation
(\ref{dollar}) connecting asymptotic {\it in} and {\it out} density
matrices. Let one suppose that there is invariance under CPT,
then there must
exist a unitary, invertible operator $\Theta$ acting on density
matrices, such that $\Theta {\overline \rho}_{in} = \rho_{out} $,
where the barred quantities denote antiparticles. Using
(\ref{dollar}), after some elementary algebraic
manipulations we obtain $ \rho_{out} =$ \$ $\rho_{in} \to $ $ \Theta
{\overline \rho}_{in} =$\$ $\Theta^{-1} {\overline \rho}_{out}$ $\to
$ ${\overline \rho}_{in} =\Theta^{-1} $\$ $\Theta^{-1} {\overline
\rho}_{out} $. But, since ${\overline \rho}_{out} =$\$${\overline
\rho}_{in} $, one arrives at ${\overline \rho}_{in} =
\Theta^{-1}$\$ $\Theta^{-1} $ \$ ${\overline \rho}_{in}$.

The latter relation, if true, would imply that \$ has an
inverse $\Theta^{-1}$ \$$\Theta^{-1} $; but this can be shown to be
impossible when one has a mixed final state, i.e., {\it decoherence} (which
is related to information loss). We omit here the details of this
last but important part, due to lack of space. The interested reader
is referred to the original literature~\cite{wald}.

From the above considerations one concludes that,
under the special circumstances described,
the generator of
CPT transformations cannot be a well-defined quantum-mechanical
operator (and thus CPT is violated at least in its strong form).
This form of violation introduces a fundamental arrow of
time/microscopic time irreversibility, unrelated in principle to CP
properties. The reader's attention is called to the fact that such
decoherence-induced CPT violation (CPTV) would occur in effective field
theories, i.e., when the low-energy experimenters do not have access to
all the degrees of freedom of QG (e.g., back-reaction
effects, \emph{etc.}). It is unknown whether full CPT invariance
could be restored in the (still elusive) complete theory of QG.

In such a case, however, there may be~\cite{wald}
a \emph{weak form of CPT invariance}, in the sense of the possible existence
of \emph{decoherence-free subspaces} in the space of states of a
matter system. If this situation is realized, then the strong form
of CPTV will not show up in any measurable quantity (that
is, scattering amplitudes, probabilities \emph{etc.}).

The weak form of CPT invariance may be stated as follows: \emph{Let}
$\psi \in {\cal H}_{\it in}$, $\phi \in {\cal H}_{\it out}$
\emph{denote pure states in the respective Hilbert spaces ${\cal H}$
of in and out states, assumed accessible to experiment. If
$\theta $ denotes the (anti-unitary) CPT operator acting on pure
state vectors, then weak CPT invariance implies the following
equality between transition probabilities}
\begin{equation}
{\cal P}(\psi \to \phi) = {\cal P}(\theta^{-1}\phi \to \theta
\psi)~. \label{weakcpt}
\end{equation}
Experimentally it is possible, at least in principle, to test
equations like (\ref{weakcpt}), in the sense that, if decoherence
occurs, it induces (among other modifications) damping factors
in the time profiles of the corresponding transition probabilities.
The diverse experimental techniques for testing decoherence
range from
terrestrial laboratory experiments (in high-energy, atomic and
nuclear physics) to astrophysical observations of light from distant
extragalactic sources and high-energy cosmic neutrinos~\cite{poland}.

In the present article, we restrict ourselves to
decoherence and CPT invariance tests within the neutral kaon
system~\cite{ehns,lopez,huet,benatti,fide}. As we argue later on,
this type of (decoherence-induced) CPTV exhibits
some fairly unique effects in $\phi$ factories~\cite{bmp},
associated with a possible modification of the
Einstein-Podolsky-Rosen (EPR) correlations of the entangled neutral
kaon states produced after the decay of the $\phi$-meson (similar
effects could be present for $B$ mesons produced in $\Upsilon$
decays).

Another possible mechanism of CPTV in QG
is the {\it spontaneous breaking of Lorentz
symmetry (SBL)}~\cite{sme}; this type of CPTV does
not necessarily imply (nor does it invoke) decoherence.
In this case the ground state of the field theoretic
system is characterized by non-trivial vacuum expectation values of
certain tensorial quantities,
\begin{equation} \langle {\cal A}_\mu \rangle \ne 0~, \quad~ {\rm
or} \quad ~\langle {\cal B}_{\mu_1\mu_2\dots}\rangle \ne 0~\quad
{\it etc.}~. \label{tensorvev}
\end{equation}
This may occur in (non-supersymmetric ground states of) string
theory and other models, such as loop QG~\cite{loops}.
Again there is an extensive literature on the subject of
experimental detection/bounding of potential Lorentz violation,
which we do not discuss here~\cite{kostelecky,ljm}. Instead we
restrict ourselves to Lorentz tests using neutral
kaons~\cite{kostelkaon}. We stress at this point that
quantum-gravitational decoherence and Lorentz violation are in
principle independent, in the sense that there exist
quantum-coherent Lorentz-violating models as well as
Lorentz-invariant decoherence scenarios~\cite{Millburn}.

The important difference between the CPTV in SBL models
and the CPTV due to the space-time foam is that
in the former case the CPT operator is well-defined, but \emph{does
not commute} with the effective Hamiltonian of the matter system. In
such cases one may parametrize the Lorentz and/or CPT breaking terms
by local field theory operators in the effective Lagrangian, leading
to a construction known as the ``standard model extension''
(SME)~\cite{sme}, which is a framework for studying precision tests of
such effects.

CPTV may also be caused due to deviations from locality, e.g.,
as advocated in \cite{lykken}, in an attempt to explain observed
neutrino `anomalies', such as the LSND result~\cite{lsnd}.
Violations of
locality could also be tested with high precision, by studying
discrete symmetries in meson systems.

If present, CPT-violating effects are expected  to be strongly suppressed, and
thus difficult to detect experimentally. Naively,
QG has a dimensionful constant, $G_N \sim
1/M_P^2$, where $M_P =10^{19}$ GeV is the Planck scale. Hence, CPT
violating and decohering effects may be expected to be suppressed by
$E^3/M_P^2 $, where $E$ is a typical energy scale of the low-energy
probe. However, there could be cases where loop resummation and other
effects in theoretical models result in much larger CPT-violating
effects, of order $\frac{E^2}{M_P}$. This happens, for instance, in
some loop gravity approaches to QG~\cite{loops}, or some
non-equilibrium stringy models of space-time foam involving open
string excitations~\cite{emn}. Such large effects may lie within the
sensitivities of current or immediate future experimental facilities
(terrestrial and astrophysical), provided that enhancements due to
the near-degeneracy take place, as in the neutral-kaon case.

When interpreting experimental results in searches for CPT
violation, one should pay particular attention to disentangling
ordinary-matter-induced effects, that mimic CPTV, from
genuine effects due to QG~\cite{poland}. The
order of magnitude of matter induced effects, especially in neutrino
experiments, is often comparable to that expected in some models of
QG, and one has to exercise caution, by carefully
examining the dependence of the alleged ``effect'' on the probe
energy, or on the oscillation length (in neutrino oscillation
experiments). In most models, but \emph{not always}, since the
QG-induced CPTV is expressed as a back-reaction effect of
matter onto space-time, it increases with the probe energy  $E$ (and
oscillation length $L$ in the appropriate situations). In contrast,
ordinary matter-induced ``fake'' CPT-violating effects
increase with $L$.

We emphasize that the phenomenology of CPTV is
complicated, and there does \emph{not} seem to be a \emph{single} figure
of merit for it. Depending on the precise way CPT might be violated
in a given model or class of models of QG, there are different ways
to test the violation~\cite{poland}. Below we describe
only a selected class of such sensitive probes of CPT symmetry and
quantum-mechanical evolution (unitarity, decoherence). We commence
the discussion by examining CPT and decoherence tests in neutral
kaon decays, and then continue with some tests at meson factories,
which are associated uniquely with a breaking of CPT in the
sense of its ill-defined nature in ``fuzzy'' decoherent space-times.
We then finish with a brief discussion of high-precision tests in
some charged kaon decays, specifically four-body $K_{\ell 4}^\pm$
decays, which have recently become very relevant, as a result of the
(significantly) increased statistics of recent
experiments~\cite{NA48}.

The structure of the article is as follows: in Section 2 we discuss
kaon tests of Lorentz symmetry within the SME
framework~\cite{kostelkaon}, and give the latest bounds and
prospects, especially from the point of view of meson
factories~\cite{adidomenico}. In Section 3 we describe tests of
decoherence-induced CPTV using (single-state) neutral kaon
systems. In Section 4 we discuss the novel EPR-like modifications in
meson factories; the latter may arise if the CPT operator is not well-defined,
as happens in some space-time foam models of QG.
We argue in favour of the unique character of such tests in
providing information on the stochastic nature of quantum
space-time, and we give some order-of-magnitude estimates within
some string-inspired  models. As we show, such models can be
falsified (or severely constrained) in next-generation (upgraded) $\phi$-meson factories, such
as DA$\Phi$NE~\cite{dafne}. The enhancement of the effect provided by the
identical decay channels $(\pi^+\pi^-, \pi^+\pi^-)$ is unique.
Finally, in Section 5 we discuss
precision tests of the discrete symmetries T, CP and CPT using a
specific type of charged kaon decays~\cite{wu,treiman}, namely $
K^{+(-)}~ \to ~\pi^+ + \pi^- + \ell (\overline \ell) + \nu_\ell
(\overline \nu_\ell) $. Recently, high statistics has been attained
by the NA48 experiment~\cite{NA48}, thereby increasing the prospects
of using such decays for precision tests of CPT symmetry.
This could be accomplished through
the study of (appropriately constructed~\cite{triple}) T-odd
observables between the $K^\pm$ modes, involving triple momentum
products of the lepton and the di-pion state $\vec p \cdot
(\vec p_1 \times \vec p_2)$, which we discuss briefly.

\section{Standard Model Extension, Lorentz Violation and Neutral
Kaons}

\subsection{Formalism and Order-of-Magnitude Estimates}

As mentioned earlier, there is a case where Lorentz symmetry is
(spontaneously) violated, in the sense of certain tensorial
quantities acquiring vacuum expectation values (\ref{tensorvev}).
Hence CPT is violated, but no quantum decoherence or unitarity loss
occurs. The generator of the CPT symmetry is a well-defined
operator, which, however, does not commute with the effective
(low-energy) Hamiltonian of the matter system.

Most microscopic models where such a violation is realized are based on
string theory with exotic (non-supersymmetric) ground states
(backgrounds)~\cite{sme}, characterized by tachyonic instabilities.
In the corresponding effective low-energy string action
tachyon fields couple to tensorial fields (gauge, \emph{etc.}),
leading to non-zero v.e.v.s of certain tensorial quantities,
thus inducing Lorentz symmetry violation in these exotic string
ground states.
Models from loop
gravity~\cite{loops} or non-commutative geometries
may also display similar types of Lorentz violation,
described by analogous terms in a SME
effective Hamiltonian.

The upshot of SME is that there is a \emph{Modified Dirac Equation}
for spinor fields $\psi$, representing leptons and quarks with
charge $q$: {\small
\begin{eqnarray} &&\left( i\gamma^\mu D^\mu - M -  a_\mu \gamma^\mu
- b_\mu \gamma_5
\gamma^\mu  - \right. \nonumber \\
&& \left. \frac{1}{2}H_{\mu\nu}\sigma ^{\mu\nu} +
ic_{\mu\nu}\gamma^\mu D^\nu + id_{\mu\nu}\gamma_5\gamma^\mu D^\nu
\right)\psi =0~, \nonumber
\end{eqnarray}}where $D_\mu = \partial_\mu - A_\mu^a T^a - qA_\mu$ is an
appropriate gauge-covariant derivative. The non-conventional terms
proportional to the coefficients $a_\mu,~  b_\mu,~ c_{\mu\nu},~
d_{\mu\nu},~ H_{\mu\nu}, \dots $, stem from corresponding local
operators of the effective Lagrangian, which are phenomenological at
this stage. The set of terms pertaining to $a_\mu~, b_\mu$ entail
CPT \& Lorentz violation, while the terms proportional to
$c_{\mu\nu}~, d_{\mu\nu}~, H_{\mu\nu}$ exhibit Lorentz violation
only.

It should be stressed that, within the SME framework (as also with
the decoherence approach to QG), CPTV does \emph{not
necessarily} imply  mass differences between particle and
antiparticles.

Some remarks are now in order, regarding the form and
order-of-magnitude estimates of the Lorentz and/or CPT violating
effects. In the approach of \cite{sme,kostelecky,kostelkaon} the SME
coefficients have been taken to be constants. Unfortunately there is
not yet a detailed microscopic model available, that would allow
for concrete predictions of the order of magnitude to be made.
Theoretically, the (dimensionful, with dimensions of energy) SME
parameters can be bounded by applying renormalization group and
naturalness assumptions to the effective local SME Hamiltonian; for example,
the bounds on $b_\mu$ so obtained are of the order of $10^{-17}~{\rm
GeV}$. At present all SME parameters should be considered as
phenomenological, to be constrained by experiment.

In general, however, the SME coefficients may not
be constant. In fact, in certain string-inspired or stochastic models of
space-time foam with Lorentz symmetry violation, the coefficients
$a_\mu, b_\mu ...$ are probe-energy ($E$) dependent, as a result of
back-reaction effects of matter onto the fluctuating space-time.

Specifically, in stochastic models of space-time foam, one may find
that on average there is no CPT and/or Lorentz violation, i.e., the
respective statistical v.e.v.s (over stochastic space-time
fluctuations)
 $\langle a_\mu~, b_\mu \rangle = 0~,$ but this is not true for
 higher order correlators of these quantities (fluctuations), i.e.,
$\langle a_\mu a_\nu \rangle \ne 0,~
 \langle b_\mu a_\nu \rangle \ne 0,~  \langle b_\mu b_\nu \rangle \ne 0~, \dots
 $. In such a case the SME effects will be
much more suppressed, since, by dimensional arguments, such
fluctuations are expected to be of order $E^4/M_P^2$, probably with
no chance of being observed in upcoming facilities, and
certainly not in neutral kaon systems in the foreseeable future.

\subsection{Tests of Lorentz Violation in Neutral Kaons}

We now turn to a brief description of experimental tests of Lorentz
symmetry within the SME framework, using neutral kaons, both
single~\cite{kostelkaon} and as entangled states at a $\phi$
factory~\cite{adidomenico}.

We begin our analysis with the single-kaon case. To determine the
relevant observable, we first recall that the wave function of the
neutral kaon, $ \Psi$, is represented as a two-component $\Psi^T =
\left(K^0,{\overline K}^0 \right)$ vector (the superscript $T$
denotes matrix transposition).

Time evolution within the rules of quantum mechanics (but with
CPT- and Lorentz-violation) is described by the equation $$\partial _t
\Psi = {\cal H} \Psi \,,$$ where the effective Hamiltonian ${\cal H}$
includes CP-violating effects, the latter being parametrized by the
conventional CP-(and T-)violating parameter of order $\epsilon_K \sim
10^{-3}$, as well as CPT-(and CP-) violating effects parametrized by the
(complex) parameter~\cite{fide} $\delta_K \sim ({\cal H}_{11} -
{\cal H}_{22})/2\Delta \lambda $, with $\Delta \lambda$ the
eigenvalue difference.

In order to isolate the terms in the SME effective Hamiltonian that
are pertinent to neutral kaon tests, one should
notice~\cite{kostelkaon} that ${\cal H}_{11} - {\cal H}_{22}$ is
flavour-diagonal, and that the parameter $\delta_K$ must be
C-violating but P,T-preserving, as a consequence of strong-interaction
properties in neutral meson evolution.

Hence one should look for terms in the SME formalism that share the
above features, namely are flavour-diagonal and violate $C$, but
preserve $T,P$. These considerations imply that $\delta_K $ is
sensitive \emph{only} to the $-a_\mu^q {\overline q} \gamma_\mu q$
quark terms in SME, where $q$ denote quark fields, with the meson
composition being denoted by $M = q_1{\overline q}_2$. The analysis
of \cite{kostelkaon}, then, leads to the following relation of the
Lorentz- and CPT-violating parameter $a_\mu$ to the CPT-violating
parameter $\delta_K$ of the neutral kaon system,
$$\delta_K \simeq i{\rm sin}\widehat{\phi} {\rm
exp}(i\widehat{\phi}) \gamma \left(\Delta a_0 - {\vec \beta_K}\cdot
\Delta {\vec a}\right)/\Delta m,$$ with the usual short-hand
notation $S$=short-lived, $L$=long-lived, $I$=interference term,
$\Delta m = m_L - m_S$, $\Delta \Gamma = \Gamma_S - \Gamma_L$,
$\widehat \phi = {\rm arc}{\rm tan}(2\Delta m / \Delta\Gamma), \quad
\Delta a_\mu \equiv a_\mu^{q_2} - a_\mu^{q_1}$, and $\beta_K^\mu =
\gamma (1, {\vec \beta}_K)$ the 4-velocity of the boosted~ kaon.

\begin{figure}[htb]
\begin{centering}
  \epsfig{file=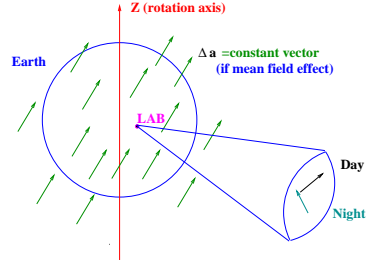, width=0.3\textwidth}
\caption{Schematic representation of searches for sidereal
variations of the CPT-violating parameter $\delta_K$ in the SME
framework. The green arrows, crossing the Earth indicate a constant
Lorentz-violating vector that characterizes the Lorentz-violating
ground state.} \label{ameas}
\end{centering}
\end{figure}

The experimental bounds on $a_\mu$ from the neutral-kaon experiments
are based on searches for sidereal variations of $\delta_K$
(day-night effects). The experimental situation is depicted
schematically in Fig.~\ref{ameas}.

From the KTeV experiment~\cite{ktev} the following bounds on the $X$
and $Y$ components of the $a_\mu$ parameter have been obtained
$$\Delta a
_X, \Delta a _Y < 9.2 \times 10^{-22}~ {\rm GeV}~,$$ where $X,Y,Z$
denote sidereal coordinates (see Fig.~\ref{ameas}).

Complementary probes of the $a_Z$ component can come from
$\phi$-factories~\cite{adidomenico}. In the case of $\phi$-factories
there is additional dependence of the CPT-violating parameter
$\delta_K$ on the polar ($\theta$) and azimuthal ($\phi$) angles
{\small
\begin{eqnarray} && \delta_{K}^\phi (|\vec p |, \theta, t) =
\frac{1}{\pi}\int_0^{2\pi}d\phi \delta_K(\vec p, t) \simeq \nonumber
\\
&& i{\rm sin}\widehat{\phi} {\rm exp}(i\widehat{\phi})
(\gamma/\Delta m) \left(\Delta a_0 + \beta_K  \Delta {a_Z}{\rm
cos}\chi {\rm cos}\theta
+ \right. \nonumber \\
&& \left. \beta_K  \Delta {a_X}{\rm sin}\chi {\rm cos}\theta {\rm
cos}(\Omega t) + \beta_K  \Delta {a_Y}{\rm sin}\chi {\rm cos}\theta
{\rm sin}(\Omega t) \right) \nonumber \end{eqnarray}}where $\Omega $
denotes  the Earth's sidereal frequency, and $\chi$ is the angle between
the laboratory Z-axis and the Earth's axis.

The experiment KLOE at DA$\Phi$NE is sensitive to $a_Z$: limits on
$\delta (\Delta a_Z)$ can be placed from forward-backward asymmetry
measurements $A_L = 2{\rm Re}\epsilon_K - 2 {\rm Re}\delta_K$. For
more details on the relevant experimental bounds we refer the reader
to the literature~\cite{adidomenico}.

We only mention at this stage that in an upgraded DA$\Phi$NE
facility, namely experiment KLOE-2 at DA$\Phi$NE-2, the expected
sensitivity is~\cite{adidomenico} $\Delta a _\mu = {\cal
O}(10^{-18})$~GeV which, however, is not competitive with the
current KTeV limits on $a_{X,Y}$ given above.

We close this subsection by pointing out that additional precision tests
can be performed using other meson factories (using B-mesons, {\it
etc.}... ), which would also allow one to test the universality of QG
Lorentz-violating effects, if observed.

\section{QG Decoherence and CPTV in Neutral Kaons}

\subsection{Stochastically Fluctuating Geometries, Light Cone
Fluctuations and Decoherence: General Ideas}

\begin{figure}[htb]
\centering
  \epsfig{file=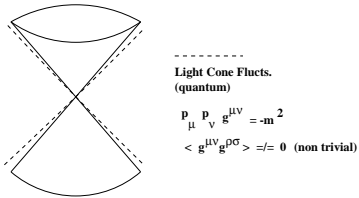, width=0.3\textwidth}
\caption{In stochastic space-time models of QG the
light cone may fluctuate, leading to decoherence and quantum
fluctuations of the speed of light in ``vacuo''.} \label{lcf}
\end{figure}

If the ground state of QG consists of ``fuzzy'' space-time, i.e.,
stochastically-fluctuating metrics, then a plethora of interesting
phenomena may occur, including light-cone
fluctuations~\cite{ford,emn} (c.f. Fig.~\ref{lcf}).
Such effects will lead to stochastic fluctuations in, say, arrival
times of photons with common energy, which can be detected with
high precision in astrophysical experiments~\cite{efmmn,ford}.
In addition, they may give rise to
decoherence of matter, in the sense of induced time-dependent
damping factors in the evolution equations of the (reduced) density
matrix of matter fields~\cite{emn,ms}.

Such ``fuzzy'' space-times are formally represented by metric
deviations which are fluctuating randomly about, say, flat Minkowski
space-time: $g_{\mu\nu} = \eta_{\mu\nu} + h_{\mu\nu}$, with $\langle
\cdots \rangle$ denoting statistical quantum averaging, and $\langle
g_{\mu\nu} \rangle = \eta_{\mu\nu} $  but $\langle h_{\mu\nu}(x)
h_{\lambda \sigma}(x') \rangle \ne 0 $,  i.e., one has only quantum
(light cone) fluctuations but not mean-field effects on dispersion
relations of matter probes. In such a situation Lorentz symmetry is
respected on the average, \emph{but} not in individual measurements.

The path of light follows null geodesics $ 0 = ds^2 =
g_{\mu\nu}dx^\mu dx^\nu $, with non-trivial fluctuations in geodesic
deviations, ${D^2n^{\mu}\over D\tau^2} =
-R^{\mu}_{\alpha\nu\beta}u^{\alpha}n^{\nu}u^{\beta}\,;$ in a
standard general-relativistic notation, $D/D\tau$ denotes the
appropriate covariant derivative operation,
$R^{\mu}_{\alpha\nu\beta}$ the (fluctuating) Riemann curvature
tensor, and $u^\mu$ ($n^\mu$) the tangential (normal) vector along
the geodesic.

Such an effect causes primarily fluctuations in the arrival time of
photons at the detector ($|\phi \rangle$=state of gravitons, $|0
\rangle$= vacuum state) {\small $$
 \Delta t_{obs}^2=|\Delta t_{\phi}^2-\Delta t_0^2 |=
{|\langle \phi| \sigma_1^2 |\phi\rangle-\langle 0| \sigma_1^2
|0\rangle|\over r^2}\equiv {|\langle \sigma_1^2 \rangle_R|\over r}\,
,
$$}
where {\small \begin{eqnarray}  && \langle \sigma_1^2 \rangle_R
=\frac{1}{8}(\Delta r)^2 \int_{r_0}^{r_1} dr \int_{r_0}^{r_1} dr'
\:\,
n^{\mu} n^{\nu} n^{\rho} n^{\sigma} \:\, \nonumber \\
&& \langle \phi| h_{\mu\nu}(x) h_{\rho\sigma}(x')+
 h_{\mu\nu}(x') h_{\rho\sigma}(x)|\phi \rangle \nonumber \end{eqnarray}}and
the two-point function of graviton fluctuations can be evaluated
using standard field theory techniques~\cite{ford}.

Apart from the stochastic metric fluctuations, however, the aforementioned
effects could also
induce decoherence of matter propagating in these types of
backgrounds~\cite{ms}, a possibility of particular interest for the
purposes of the present article.
Through the theorem of Wald~\cite{wald}, this implies that
the CPT operator is not well-defined, and hence one also has a
breaking of CPT symmetry.

We now proceed to describe briefly the general formalism used
for parametrizing such QG-induced decoherence, as far as the
CPT-violating effects on matter are concerned.

\subsection{Formalism for the Phenomenology of QG-induced Decoherence}

In this subsection we shall be very brief, giving
the reader a flavor of the formalism underlying such decoherent systems.
We shall discuss first a model-independent parametrization of
decoherence, applicable not only to QG media, but
covering a more general situation.

If the effects of the environment are such
that the modified evolution equation of the (reduced) density matrix
of matter $\rho$~\cite{kiefer} is linear, one can write
down a Lindblad evolution equation~\cite{lindblad}, provided that
(i) there is (complete) positivity of $\rho$, so that negative
probabilities do not arise at any stage of the evolution, (ii) the
energy of the matter system is conserved on the average, and (iii)
the entropy is increasing monotonically.

For $N$-level systems, the generic decohering Lindblad evolution for $\rho$ reads
\begin{eqnarray}
&&\frac{\partial \rho_\mu}{\partial t} = \sum_{ij} h_i\rho_j
{f}_{ij\mu} + \sum_{\nu} {L}_{\mu\nu}\rho_\nu~, \nonumber
\\
&& \quad~\mu, \nu = 0, \dots N^2 -1, \quad i,j = 1, \dots N^2 -1~,
\nonumber
\end{eqnarray}
where the $h_i$ are Hamiltonian terms, expanded in an appropriate
basis, and the decoherence matrix $L$ has the form:
${L}_{0\mu}={L}_{\mu 0} =0~,$ ${L}_{ij} = \frac{1}{4}\sum_{k,\ell
,m} c_{l\ell}\left(-f_{i\ell m}f_{kmj} + f_{k i m}f_{\ell m
j}\right)~,$ with $c_{ij}$ a positive-definite matrix and $f_{ijk}$
the structure constants of the appropriate $SU(N)$ group.
In this generic phenomenological description of decoherence, the
elements ${L}_{\mu\nu}$ are free parameters,
to be determined by experiment. We shall come back to this point in
the next subsection, where we discuss neutral kaon decays.

A rather characteristic feature of this equation is the
appearance of exponential damping, $e^{-(...)t}$, in interference terms
of the pertinent quantities (for instance, matrix elements $\rho$,
or asymmetries in the case of the kaon system, see below).
The exponents are proportional to (linear
combinations) of the elements of the decoherence
matrix~\cite{lindblad,ehns,kiefer}.
Note, however, that Lindblad type evolution is
\emph{not} the most generic evolution for QG models. In
cases of space-time foam corresponding to {\it stochastically
(random) fluctuating space-times}, such as the situations causing
light-cone fluctuations examined previously, there is a different
kind of decoherent evolution, with damping that is quadratic in
time, i.e., one has a $e^{-(...)t^2}$ suppression of interference
terms in the relevant observables.

A  specific  model  of  stochastic  space-time  foam  is  based  on  a
particular kind of gravitational foam~\cite{emn,emw,ms}, consisting of
``real''   (as   opposed  to   ``virtual'')   space-time  defects   in
higher-dimensional  space   times,  in  accordance   with  the  modern
viewpoint of our  world as a brane hyper-surface  embedded in the bulk
space-time~\cite{polch}.   This   model  is  quite   generic  in  some
respects, and we will use it  later to estimate the order of magnitude
of novel CPT violating effects in entangled states of kaons.

A model of space-time foam \cite{emw} can be based on a number
(determined by target-space supersymmetry) of parallel brane worlds
with three large spatial dimensions. These brane worlds
move in a bulk space-time,
containing a ``gas'' of point-like bulk branes, termed
``D-particles'', which are stringy space-time solitonic defects. One
of these branes is the observable Universe. For an observer on the
brane the crossing D-particles will appear as twinkling space-time
defects, i.e. microscopic space-time fluctuations. This will give
the four-dimensional brane world a ``D-foamy'' structure. Following
work on gravitational decoherence \cite{emn,ms}, the target-space
metric state, which is close to being flat, can be represented
schematically as a density matrix
\begin{equation}
\rho_{\mathrm{grav}}=\int d\,^{5}r\,\,f\left(  r_{\mu}\right)
\left| g\left(  r_{\mu}\right)  \right\rangle \left\langle g\left(
r_{\mu}\,\right)\right|  .\, \label{gravdensity}%
\end{equation}
\bigskip The parameters $r_{\mu}\,\left(  \mu=0,1 \ldots \right)  $ pertain
to appropriate space-time metric deformations and are
stochastic, with a Gaussian distribution $\,f\left(  r_{\mu}\,\right)
$
characterized by the averages%
\[
\left\langle r_{\mu}\right\rangle =0,\;\left\langle r_{\mu}r_{\nu
}\right\rangle =\Delta_{\mu}\delta_{\mu\nu}\,.
\]
This model will be studied in more detail in section 4.

We will assume that
the fluctuations of the metric felt by two entangled neutral mesons
are independent, and $\Delta_{\mu}\sim O\left(  \frac{E^{2}}%
{M_{P}^{2}}\right)  $, i.e., very small. As matter moves through the
space-time foam in a typical ergodic picture, the effect of time
averaging is assumed to be equivalent to an ensemble average.
For our present discussion we consider a
semi-classical picture for the metric, and therefore $\left|  g\left(
r_{\mu}\right)  \right\rangle $ in (\ref{gravdensity}) is a
coherent state.

In the specific model of foam discussed in \cite{ms}, there is a
recoil effect of the D-particle, as a result of its scattering with
stringy excitations that live on the brane world and represent
low-energy ordinary matter. As the space-time defects, propagating
in the bulk space-time, cross the brane hyper-surface from the bulk
in random directions, they scatter with matter. The associated
distortion of space-time caused by this scattering can be considered
dominant only along the direction of motion of the matter probe.
Random fluctuations are then considered about an average flat
Minkowski space-time. The result is an effectively  two-dimensional
approximate fluctuating metric describing the main effects~\cite{ms}

{\scriptsize\begin{eqnarray}
&& g^{\mu\nu}= \nonumber \\
&& \left(\begin{array}{cc}
  -(a_1+1)^2 + a_2^2 & -a_3(a_1+1) +a_2(a_4+1) \\
  -a_3(a_1+1) +a_2(a_4+1) & -a_3^2+(a_4+1)^2 \\
\end{array}\right). \nonumber \\
\label{flct}
 \end{eqnarray}}
The $a_i$ represent the fluctuations and are assumed to be random
variables, satisfying $\langle a_i\rangle =0$ and  $\langle a_i a_j\rangle =
\delta_{ij}\sigma_i$.

Such a (microscopic) model of space-time foam is not of Lindblad
type, as can be seen~\cite{ms} by considering the oscillation
probability for, say, two-level scalar systems describing
oscillating neutral kaons, $K^0 ~\leftrightarrow \overline{K}^0$.
In the approximation of small fluctuations one
finds the following form for the oscillation probability of the
two-level scalar system:
\begin{eqnarray}
&& \langle e^{i(\omega _{1}-\omega _{2})t}\rangle = \nonumber \\
&&\frac{4\tilde{d}^{2}}{(P_{1}P_{2})^{1/2}}\exp \left( \frac{\chi _{1}}{%
\chi _{2}}\right) \exp (i\tilde{b}t)\,, \nonumber \label{timedep}
\end{eqnarray}%
where $\omega_i,~i=1,2$ are the appropriate energy levels~\cite{ms}
of the two-level kaon system in the background of the fluctuating
space-time~(\ref{flct}), and

{\small \begin{eqnarray*}
\chi _{1} &=&-4(\tilde{d}^{2}\sigma _{1}+\sigma _{4}k^{4})\tilde{b}%
^{2}t^{2}+2i\tilde{d}^{2}\widetilde{b}^{2}\widetilde{c}k^{2}\sigma
_{1}\sigma _{4}t^{3}, \\
\chi _{2} &=&4\tilde{d}^{2}-2i\tilde{d}^{2}(k^{2}\tilde{c}\sigma _{4}+2%
\tilde{b}\sigma _{1})t+ \nonumber \\
&& \widetilde{b}k^{2}\left( \widetilde{b}k^{2}-2%
\widetilde{d}^{2}\widetilde{c}\right) \sigma _{1}\sigma _{4}, \\
P_{1} &=&4\tilde{d}^{2}+2i\widetilde{d}\widetilde{b}\left( k^{2}-\widetilde{d%
}\right) \sigma _{2}t+\tilde{b}^{2}k^{4}\sigma _{2}\sigma _{3}t^{2}, \\
P_{2} &=&4\tilde{d}^{2}-2i\widetilde{d}^{2}\left(
k^{2}\widetilde{c}\sigma _{4}+2\widetilde{b}\sigma _{1}\right)
t+ \mathcal{O}\left( \sigma ^{2}\right)~,
\end{eqnarray*}}
with {\small \begin{eqnarray*}
\begin{array}{c}
\tilde{b}=\sqrt{k^{2}+m_{1}^{2}}-\sqrt{k^{2}+m_{2}^{2}}, \nonumber \\
\tilde{c}%
=m_{1}^{2}(k^{2}+m_{1}^{2})^{-3/2}-m_{2}^{2}(k^{2}+m_{2}^{2})^{-3/2}, \nonumber \\
\tilde{d}=\sqrt{k^{2}+m_{1}^{2}}\sqrt{k^{2}+m_{2}^{2}}.\nonumber
\end{array}
\end{eqnarray*}}
From this expression one can see~\cite{ms} that the stochastic
model of space-time foam leads to a modification of oscillation
behavior quite distinct from that of the Lindblad formulation. In
particular,
the transition probability displays a Gaussian time-dependence,
decaying as $e^{-(...)t^2}$, a modification of
the oscillation period, as well as additional power-law fall-off.

{}From this characteristic time-dependence, one can obtain bounds
for the fluctuation strength of space-time foam in kaon systems. In
the context of this presentation, we restrict ourselves to
Lindblad decoherence tests using only neutral kaons. However, when
discussing the CPTV effects of foam on entangled states we make use
of this specific model of stochastically fluctuating D-particle
foam~\cite{emw,ms}, in order to demonstrate the effects explicitly
and obtain definite order-of-magnitude estimates~\cite{bms}.

\subsection{Experiments involving Single-Kaon States}

As mentioned in the previous subsection, QG may induce decoherence
and oscillations $K^0 \leftrightarrow {\overline
K}^0$~\cite{ehns,lopez}, thereby implying a two-level quantum
mechanical system interacting with a QG ``environment''. Adopting
the general assumptions of average energy conservation and monotonic
entropy increase, the simplest model for parametrizing decoherence
(in a rather model-independent way) is the (linear) Lindblad
approach mentioned earlier. Not all entries of a general
decoherence matrix are physical, and in order to isolate the
physically relevant entries one must invoke specific
assumptions, related to the symmetries of the particle system in
question. For the neutral kaon system, such an extra assumptions
is that the QG medium respects
the $\Delta S=\Delta Q$ rule.
In such a case, the modified Lindblad evolution equation
(\ref{evoleq}) for
the respective density matrices of neutral kaon matter can be
parametrized as follows~\cite{ehns}:
$$\partial_t \rho = i[\rho, H] + \delta\H \rho~,$$
where {\small $$H_{\alpha\beta}=\left( \begin{array}{cccc}  - \Gamma
& -\coeff{1}{2}\delta \Gamma
& -{\rm Im} \Gamma _{12} & -{\rm Re}\Gamma _{12} \\
 - \coeff{1}{2}\delta \Gamma
  & -\Gamma & - 2{\rm Re}M_{12}&  -2{\rm Im} M_{12} \\
 - {\rm Im} \Gamma_{12} &  2{\rm Re}M_{12} & -\Gamma & -\delta M    \\
 -{\rm Re}\Gamma _{12} & -2{\rm Im} M_{12} & \delta M   & -\Gamma
\end{array}\right) $$} and
$$ {\delta\H}_{\alpha\beta} =\left( \begin{array}{cccc}
 0  &  0 & 0 & 0 \\
 0  &  0 & 0 & 0 \\
 0  &  0 & -2\alpha  & -2\beta \\
 0  &  0 & -2\beta & -2\gamma \end{array}\right)~.$$
Positivity of $\rho$ requires: $\alpha, \gamma  > 0,\quad
\alpha\gamma>\beta^2$. Notice that $\alpha,\beta,\gamma$ violate
{\it both}  CPT, due to their decohering nature~\cite{wald}, and CP
symmetry, as they do not commute with the CP operator
$\widehat{CP}$~\cite{lopez}: $\widehat{CP} = \sigma_3 \cos\theta +
\sigma_2 \sin\theta$,$~~~~~[\delta\H_{\alpha\beta}, \widehat{CP} ]
\ne 0$.

An important remark is now in order. As pointed out in
\cite{benatti}, although the above parametrization is sufficient for
a single-kaon state to have a positive definite density matrix (and
hence probabilities) this is \emph{not} true when one considers
the evolution of entangled kaon states ($\phi$-factories).
In this latter case,
complete positivity is guaranteed only if
the further conditions
\begin{equation}\label{cons}
\alpha = \gamma~ {\rm and} ~\beta = 0
\end{equation}
are imposed. When incorporating entangled states, one should either
consider possible new effects (such as the $\omega$-effect
considered below) or apply the constraints (\ref{cons}) also to
single kaon states~\cite{benatti}. This is not necessarily the case
when other non-entangled particle states, such as neutrinos, are
considered, in which case the $\alpha,\beta,\gamma$ parametrization
of decoherence may be applied. Experimentally the complete
positivity hypothesis can be tested explicitly. In what follows, as
far as single-kaon states are concerned, we keep the
$\alpha,\beta,\gamma$ parametrization, and give the available
experimental bounds for them, but we always have in mind the
constraint (\ref{cons}) when referring to entangled kaon states in a
$\phi$-factory.

As already mentioned, when testing CPT symmetry with neutral kaons
one should be careful to distinguish two types of CPTV: {\bf (i)} CPTV
within Quantum Mechanics~\cite{fide}, leading to possible
differences between particle-antiparticle masses and widths: $\delta
m= m_{K^0} - m_{{\overline K}^0}$, $\delta \Gamma = \Gamma_{K^0}-
\Gamma_{{\overline K}^0} $. This type of CPTV could be,
for instance,
due to (spontaneous) Lorentz violation~\cite{sme}. In that
case the CPT operator is well-defined as a quantum mechanical
operator, but does not commute with the Hamiltonian of the system.
This, in turn, may lead to mass and width differences between particles
and antiparticles, among other effects. {\bf (ii)} CPTV through
decoherence~\cite{ehns,poland} via the parameters
$\alpha,\beta,\gamma$ (entanglement with the QG ``environment'',
leading to modified evolution for $\rho$ and $\$ \ne S~S^\dagger $).
In the latter case the CPT operator may not be well-defined, which
implies novel effects when one uses entangled states of kaons, as we
shall discuss in the next subsection.

\begin{figure}[htb]
\centering
  \epsfig{file=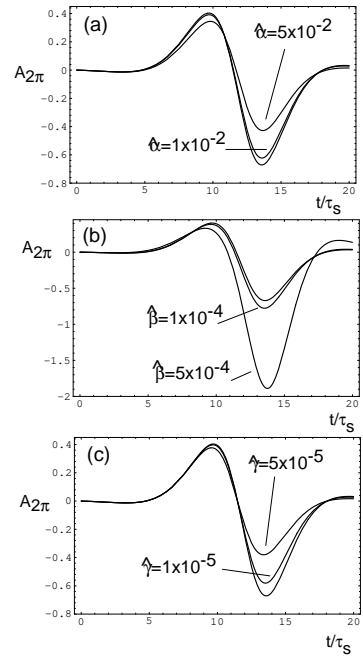, width=0.3\textwidth} \hfill
\caption{Neutral kaon decay asymmetries $A_{2\pi}$~\cite{lopez}
indicating the effects of QG-induced decoherence.}
\label{AT}
\end{figure}

\begin{figure}[htb]
\centering
\epsfig{file=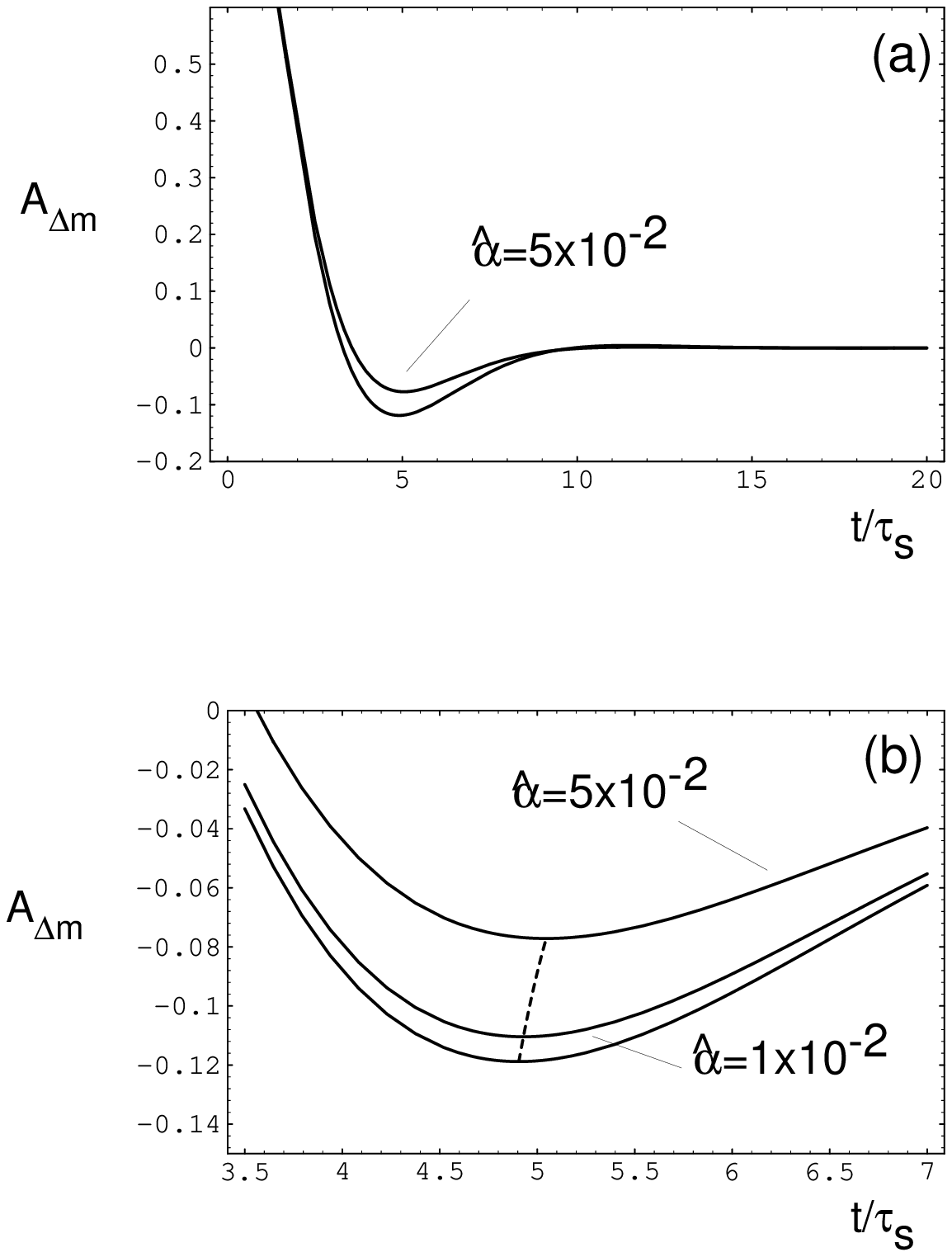, width=0.3\textwidth} \hfill
\caption{Typical neutral kaon decay asymmetries $A_{\Delta
m}$~\cite{lopez} indicating the effects of QG-induced
decoherence.} \label{AT2}
\end{figure}

\begin{figure}[htb]
\centering
  \epsfig{file=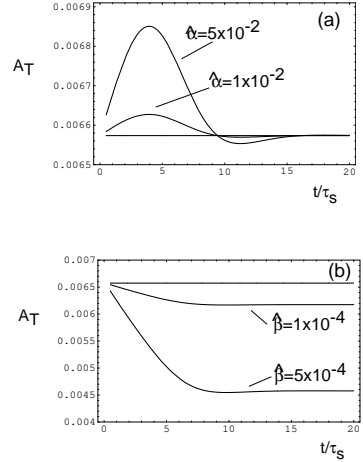, width=0.3\textwidth}
\caption{Typical neutral kaon decay asymmetries $A_T$~\cite{lopez}
indicating the effects of QG-induced decoherence.}
\label{AT3}
\end{figure}

\begin{table}[thb]
\begin{center}
\begin{tabular}{lcc}
\underline{Process}&QMV&QM\\
$A_{2\pi}$&$\not=$&$\not=$\\
$A_{3\pi}$&$\not=$&$\not=$\\
$A_{\rm T}$&$\not=$&$=$\\
$A_{\rm CPT}$&$=$&$\not=$\\
$A_{\Delta m}$&$\not=$&$=$\\
$\zeta$&$\not=$&$=$
\end{tabular}
\caption{Qualitative comparison of predictions for various
observables in CPT-violating theories beyond (QMV) and within (QM)
quantum mechanics. Predictions either differ ($\not=$) or agree
($=$) with the results obtained in conventional quantum-mechanical
CP violation. Note that these frameworks can be qualitatively
distinguished via their predictions for $A_{\rm T}$, $A_{\rm CPT}$,
$A_{\Delta m}$, and $\zeta$.} \label{Table2}
\end{center}
\hrule
\end{table}

The important point to notice is that the two types of CPTV can be
{\it disentangled experimentally}~\cite{lopez}. The relevant
observables are defined as $ \VEV{O_i}= {\rm Tr}\,[O_i\rho] $. For
neutral kaons, one looks at decay asymmetries for $K^0, {\overline
K}^0$, defined as:
$$A (t) = \frac{
    R({\bar K}^0_{t=0} \rightarrow
{\bar f} ) -
    R(K^0_{t=0} \rightarrow
f ) } { R({\bar K}^0_{t=0} \rightarrow {\bar f} ) +
    R(K^0_{t=0} \rightarrow
f ) }~,$$ where $R(K^0\rightarrow f) \equiv \Tr[O_{f}\rho (t)]=$
denotes the decay rate into the final state $f$ (starting from a
pure $ K^0$ state at $t=0$).

In the case of neutral kaons, one may consider the following set of
asymmetries: (i) {\it identical final states}: $f={\bar f} = 2\pi $:
$A_{2\pi}~,~A_{3\pi}$, (ii) {\it semileptonic} : $A_T$ (final states
$f=\pi^+l^-\bar\nu\ \not=\ \bar f=\pi^-l^+\nu$), $A_{CPT}$
(${\overline f}=\pi^+l^-\bar\nu ,~ f=\pi^-l^+\nu$), $A_{\Delta m}$.
Typically, for instance when final states are $2\pi$, one has  a
time evolution of the decay rate $R_{2\pi}$: $ R_{2\pi}(t)=c_S\,
e^{-\Gamma_S t}+c_L\, e^{-\Gamma_L t} + 2c_I\, e^{-\Gamma
t}\cos(\Delta mt-\phi)$, where $S$=short-lived, $L$=long-lived,
$I$=interference term, $\Delta m = m_L - m_S$, $\Delta \Gamma =
\Gamma_S - \Gamma_L$, $\Gamma =\frac{1}{2}(\Gamma_S + \Gamma_L)$.
One may define the {\it decoherence parameter}
$\zeta=1-{c_I\over\sqrt{c_Sc_L}}$, as a (phenomenological) measure
of quantum decoherence induced in the system~\cite{fide}. For larger
sensitivities one can look at this parameter in the presence of a
regenerator~\cite{lopez}. In our decoherence scenario, $\zeta$
corresponds to a particular combination of the decoherence
parameters~\cite{lopez}:
$$ \zeta \to \frac{\widehat \gamma}{2|\epsilon ^2|} -
2\frac{{\widehat \beta}}{|\epsilon|}{\rm sin} \phi~,$$ with the
notation $\widehat{\gamma} =\gamma/\Delta \Gamma $, \emph{etc}.
Hence, ignoring the constraint (\ref{cons}), the best bounds on
$\beta$, or -turning the logic around- the most sensitive tests of
complete positivity in kaons, can be placed by implementing a
regenerator~\cite{lopez}.

The experimental tests (decay asymmetries) that can be performed in
order to disentangle decoherence from quantum-mechanical CPT
violating effects are summarized in Table \ref{Table2}. In Figures
\ref{AT}, \ref{AT2}, \ref{AT3} we give typical profiles of several
decay asymmetries~\cite{lopez}, from where bounds on QG decohering
parameters can be extracted. At present there are experimenatl bounds
available from CPLEAR measurements~\cite{cplear} $\alpha
< 4.0 \times 10^{-17} ~{\rm GeV}~, ~|\beta | < 2.3. \times 10^{-19}
~{\rm GeV}~, ~\gamma < 3.7 \times 10^{-21} ~{\rm GeV} $, which are
not much different from theoretically expected values in some
optimistic scenarios~\cite{lopez} $\alpha~,\beta~,\gamma = O(\xi
\frac{E^2}{M_{P}})$.

Recently, the experiment KLOE at DA$\Phi$NE updated these limits by
measuring for the first time the $\gamma$ decoherence parameter for
entangled kaon states~\cite{adidomenico}, as well as the (naive)
decoherence parameter $\zeta$ (to be specific, the KLOE
Collaboration has presented measurements for two $\zeta$ parameters,
one, $\zeta_{LS}$, pertaining to an expansion in terms of $K_L,K_S$
states, and the other, $\zeta_{0\bar 0}$, for an expansion in terms
of $K^0,\overline K^0$  states). We remind the reader once more
that, under the assumption of complete positivity for entangled
meson states~\cite{benatti}, theoretically there is only one
parameter to parametrize Lindblad decoherence, since $\alpha =
\gamma$, $\beta = 0$. In fact, the KLOE experiment has the greatest
sensitivity to this parameter $\gamma$. The latest KLOE
measurement~\cite{adidomenico} for $\gamma$ yields $\gamma_{\rm
KLOE} = (1.1^{+2.9}_{-2.4} \pm 0.4) \times 10^{-21}~{\rm GeV}$, i.e.
$\gamma < 6.4 \times 10^{-21}~{\rm GeV}$, competitive with the
corresponding CPLEAR bound~\cite{cplear} discussed above. It is
expected that this bound could be improved by an order of magnitude
in upgraded facilities, such as KLOE-2 at
DA$\Phi$NE-2~\cite{adidomenico}, where one expects $\gamma_{\rm
upgrade} \to \pm 0.2 \times 10^{-21} ~{\rm GeV}$.

The reader should also bear in mind that the Lindblad linear
decoherence is not the only possibility for a parametrization of QG
effects, see for instance the stochastically fluctuating space-time
metric approach discussed in Section 3.1 above. Thus, direct tests of
the complete positivity hypothesis in entangled states, and hence
the theoretical framework {\it per se}, should be performed by
independent measurements of all the three decoherence parameters
$\alpha,\beta,\gamma$; as far as we understand~\footnote{We thank A.
Di Domenico for informative discussions on this point.}, such data
are currently available in kaon factories, but not yet analyzed in
detail~\cite{adidomenico}.

\section{CPTV and Modified EPR Correlations
of Entangled Neutral Kaon States}

\subsection{EPR Correlations in Particle Physics}

We now come to a description of an entirely novel effect~\cite{bmp}
of CPTV due to the ill-defined nature of the CPT
operator, which is \emph{exclusive} to neutral-meson factories, for
reasons explained below. The effects are qualitatively similar
for kaon and $B$-meson factories~\cite{bomega}, with the
important observation that in kaon factories there is a particularly
good channel, that of both correlated kaons decaying to $\pi^+\pi^-$.
In that channel the sensitivity of the effect increases because
the complex parameter $\omega$, parametrizing the relevant
EPR modifications~\cite{bmp}, appears in the particular
combination $|\omega|/|\eta_{+-}|$, with
$|\eta_{+-}| \sim 10^{-3}$. In the case of  $B$-meson factories
one should focus instead on the ``same-sign'' di-lepton
channel~\cite{bomega}, where high statistics is expected.

In this article we restrict ourselves to the case of
$\phi$-factories, referring the interested reader to the
literature~\cite{bomega} for the $B$-meson applications. We commence
our discussion by briefly reminding the reader of
EPR particle correlations.

The EPR effect was originally proposed as a {\it paradox}, testing the
foundations of Quantum Theory. There was the question whether
quantum correlations between spatially separated events implied
instant transport of information that would contradict special relativity.
It was eventually realized that no super-luminal propagation was
actually involved in the EPR phenomenon, and thus there was no
conflict with relativity.

\begin{figure}
\centering
  \epsfig{file=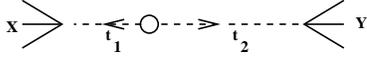, width=0.3\textwidth}
\caption{Schematic representation of the decay of a $\phi$-meson at
rest (for definiteness) into pairs of entangled neutral kaons, which
eventually decay on the two sides of the detector.}
\label{epr}\end{figure}

The EPR effect has been confirmed experimentally, e.g., in meson
factories: (i) a pair of particles can be created in a definite
quantum state, (ii) move apart and, (iii) eventually decay when they
are widely (spatially) separated (see Fig.~\ref{epr} for a schematic
representation of an EPR effect in a meson factory). Upon making a
measurement on one side of the detector and identifying the decay products,
we \emph{infer} the type of products appearing on the
other side; this is essentially the EPR correlation phenomenon.
It does \emph{not} involve any \emph{simultaneous measurement} on
both sides, and hence there is no contradiction with special
relativity. As emphasized by Lipkin~\cite{lipkin}, the EPR
correlations between different decay modes should be taken into
account when interpreting any experiment.

\subsection{CPTV and Modified EPR-Correlations in $\phi$ Factories:
the $\omega$-Effect}

In the case of $\phi$ factories
it was \emph{claimed }~\cite{dunietz} that
due to EPR correlations,  \emph{irrespective} of CP, and CPT
violation, the \emph{final state} in $\phi$ decays: $e^+ e^-
\Rightarrow \phi \Rightarrow K_S K_L $ always contains $K_LK_S$
products.
This is a direct consequence
of imposing the requirement of {\it Bose statistics}
on the state $K^0 {\overline K}^0$ (to which the $\phi$ decays);
this, in turn, implies that the physical neutral meson-antimeson
state must be {\it symmetric} under C${\cal P}$, with C the charge
conjugation and ${\cal P}$ the operator that permutes the spatial
coordinates. Assuming {\it conservation} of angular momentum, and a
proper existence of the {\it antiparticle state} (denoted by a bar),
one observes that: for $K^0{\overline K}^0$ states which are
C-conjugates with C$=(-1)^\ell$ (with $\ell$ the angular momentum
quantum number), the system has to be an eigenstate of the
permutation operator ${\cal P}$ with eigenvalue $(-1)^\ell$. Thus,
for $\ell =1$: C$=-$ $\rightarrow {\cal P}=-$.  Bose statistics
ensures that for $\ell = 1$ the state of two \emph{identical} bosons
is \emph{forbidden}. Hence, the  initial entangled state:

{\scriptsize\begin{eqnarray*} &&|i> = \frac{1}{\sqrt{2}}\left(|K^0({\vec
k}),{\overline K}^0(-{\vec k})>
- |{\overline K}^0({\vec k}),{K}^0(-{\vec k})>\right)  \nonumber \\
&& = {\cal N} \left(|K_S({\vec k}),K_L(-{\vec k})> - |K_L({\vec
k}),K_S(-{\vec k})> \right)\nonumber
\end{eqnarray*}}with the normalization factor ${\cal N}=\frac{\sqrt{(1
+ |\epsilon_1|^2) (1 + |\epsilon_2|^2
)}}{\sqrt{2}(1-\epsilon_1\epsilon_2)} \simeq \frac{1 +
|\epsilon^2|}{\sqrt{2}(1 - \epsilon^2)}$, and
$K_S=\frac{1}{\sqrt{1 + |\epsilon_1^2|}}\left(|K_+> + \epsilon_1
|K_->\right)$, $K_L=\frac{1}{\sqrt{1 + |\epsilon_2^2|}}\left(|K_->
+ \epsilon_2 |K_+>\right)$, where $\epsilon_1, \epsilon_2$ are
complex parameters, such that $\epsilon \equiv \epsilon_1 +
\epsilon_2$ denotes the CP- \& T-violating parameter, whilst $\delta
\equiv \epsilon_1 - \epsilon_2$  parametrizes the CPT \& CP violation
within quantum mechanics~\cite{fide}, as discussed previously.
The $K^0 \leftrightarrow {\overline K}^0$ or $K_S
\leftrightarrow K_L$ correlations are apparent after evolution, at
any time $t > 0$ (with $t=0$ taken as the moment of the $\phi$ decay).

In the above considerations there is an implicit assumption,
which was noted in \cite{bmp}. The above arguments are valid
independently of CPTV, provided such violation occurs
within quantum mechanics, e.g., due to spontaneous Lorentz
violation, where the CPT operator is well defined.

If, however, CPT is \emph{intrinsically} violated, due, for instance,
to decoherence scenarios in space-time foam, then
the factorizability property of the
super-scattering matrix \$ breaks down, \$ $\ne SS^\dagger $, and
the generator of CPT is not well defined~\cite{wald}. Thus, the
concept of an ``antiparticle'' may be \emph{modified} perturbatively! The
perturbative modification of the properties of the antiparticle is
important, since the antiparticle state is a physical state which
exists, despite the ill-definition of the CPT operator. However, the
antiparticle Hilbert space will have components that are
\emph{independent} of the particle Hilbert
space.

In such a case,
the neutral mesons $K^0$ and ${\overline K}^0$ should \emph{no
longer} be treated as \emph{indistinguishable particles}. As a
consequence~\cite{bmp}, the initial entangled state in $\phi$
factories $|i>$, after the $\phi$-meson decay, will acquire a component
with opposite permutation (${\cal P}$) symmetry:

{\scriptsize \begin{eqnarray*} |i> &=& \frac{1}{\sqrt{2}}\left(|K_0({\vec
k}),{\overline K}_0(-{\vec k})>
- |{\overline K}_0({\vec k}),K_0(-{\vec k})> \right)\nonumber \\
&+&  \frac{\omega}{2} \left(|K_0({\vec k}), {\overline K}_0(-{\vec k})> + |{\overline K}_0({\vec
k}),K_0(-{\vec k})> \right)  \bigg]  \nonumber \\
& = & \bigg[ {\cal N} \left(|K_S({\vec
k}),K_L(-{\vec k})>
- |K_L({\vec k}),K_S(-{\vec k})> \right)\nonumber \\
&+&  \omega \left(|K_S({\vec k}), K_S(-{\vec k})> - |K_L({\vec
k}),K_L(-{\vec k})> \right)  \bigg]~, \nonumber
\end{eqnarray*}}where ${\cal N}$ is an appropriate normalization factor,
and $\omega = |\omega |e^{i\Omega}$ is a complex parameter,
parametrizing the intrinsic CPTV modifications of the EPR
correlations. Notice that, as a result of the $\omega$-terms, there
exist, in the two-kaon state,
$K_SK_S$ or $K_LK_L$ combinations,
which
entail important effects to the various decay channels. Due to this
effect, termed the $\omega$-effect by the authors of \cite{bmp},
there is \emph{contamination} of ${\cal P}$(odd) state with ${\cal P}$({\rm even})
terms. The $\omega$-parameter controls the amount of contamination
of the final ${\cal P}$(odd) state by the ``wrong'' (${\cal P}$(even)) symmetry state.

Later in this section  we will present a microscopic model where such a
situation is realized explicitly. Specifically,
an $\omega$-like effect appears due to the evolution in the space-time
foam, and the corresponding parameter turns out to be
purely imaginary and time-dependent~\cite{bms}.

\subsection{$\omega$-Effect Observables}

To construct the appropriate observable for the possible detection
of the $\omega$-effect, we consider the $\phi$-decay amplitude
depicted in Fig.~\ref{epr}, where one of the kaon products decays to
the  final state $X$ at $t_1$ and the other to the final state $Y$
at time $t_2$. We take $t=0$ as the moment of the $\phi$-meson
decay.

The relevant amplitudes read:
\begin{eqnarray*}
A(X,Y) = \langle X|K_S\rangle \langle Y|K_S \rangle \, {\cal N}
\,\left( A_1  +  A_2 \right)~, \nonumber
\end{eqnarray*}
with \begin{eqnarray*}
 A_1  &=& e^{-i(\lambda_L+\lambda_S)t/2}
[\eta_X  e^{-i \Delta\lambda \Delta t/2}
-\eta_Y  e^{i \Delta\lambda \Delta t/2}]\nonumber \\
A_2  &=&  \omega [ e^{-i \lambda_S t} - \eta_X \eta_Y e^{-i
\lambda_L t}] \nonumber
\end{eqnarray*}
denoting the CPT-allowed and CPT-violating parameters respectively,
and $\eta_X = \langle X|K_L\rangle/\langle X|K_S\rangle$ and $\eta_Y
=\langle Y|K_L\rangle/\langle Y|K_S\rangle$. In the above formulae, $t$
is the sum of the decay times $t_1, t_2$ and $\Delta t $ is their
difference (assumed positive).

\begin{figure}[htb]
\centering
  \epsfig{file=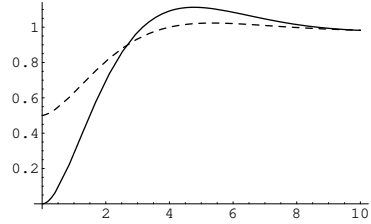, width=0.3\textwidth}
\caption{A characteristic case of the intensity $I(\Delta t)$, with
$|\omega|=0$ (solid line)  vs  $I(\Delta t)$ (dashed line) with
$|\omega|=|\eta_{+-}|$, $\Omega = \phi_{+-} - 0.16\pi$, for
definiteness~\cite{bmp}.} \label{intensomega}\end{figure}

The ``intensity'' $I(\Delta t)$ is the
desired \emph{observable} for a detection of the $\omega$-effect,
\begin{eqnarray*} I (\Delta t) \equiv \frac{1}{2} \int_{\Delta
t}^\infty dt\, |A(X,Y)|^2~. \nonumber
\end{eqnarray*}
depending only on $\Delta t$.

Its time profile reads~\cite{bmp}:

{\scriptsize \begin{eqnarray*} &&
I (\Delta t) \equiv \frac{1}{2} \int_{|\Delta t|}^\infty dt\,
|A(\pi^+\pi^-,\pi^+\pi^-)|^2  = \nonumber
\\ && |\langle\pi^{+}\pi^{-}|K_S\rangle|^4 |{\cal N}|^2 |\eta_{+-}|^2
\bigg[ I_1  + I_2  +  I_{12} \bigg]~,\nonumber
\end{eqnarray*}}where

{\scriptsize \begin{eqnarray*} && I_1 (\Delta t) =
\frac{e^{-\Gamma_S \Delta t} + e^{-\Gamma_L \Delta t} - 2
e^{-(\Gamma_S+\Gamma_L) \Delta t/2} \cos(\Delta m \Delta t)}
{\Gamma_L+\Gamma_S}
\nonumber \\
&& I_2 (\Delta t) =  \frac{|\omega|^2 }{|\eta_{+-}|^2}
\frac{e^{-\Gamma_S \Delta t} }{2 \Gamma_S}
\nonumber \\
&& I_{12} (\Delta t) = - \frac{4}{4 (\Delta m)^2 + (3 \Gamma_S +
\Gamma_L)^2}  \frac{|\omega|}{|\eta_{+-}|} \times
\nonumber \\
&&\bigg[ 2 \Delta m \bigg( e^{-\Gamma_S \Delta t} \sin(\phi_{+-}-
\Omega) - \nonumber \\
&&  e^{-(\Gamma_S+\Gamma_L) \Delta t/2} \sin(\phi_{+-}- \Omega
+\Delta m \Delta t)\bigg)
\nonumber \\
&&  - (3 \Gamma_S + \Gamma_L) \bigg(e^{-\Gamma_S \Delta t}
\cos(\phi_{+-}- \Omega) - \nonumber \\
&& e^{-(\Gamma_S+\Gamma_L) \Delta t/2} \cos(\phi_{+-}- \Omega
+\Delta m \Delta t)\bigg)\bigg]~, \nonumber
\end{eqnarray*}}with $\Delta m = m_S - m_L$ and $\eta_{+-}= |\eta_{+-}|
e^{i\phi_{+-}}$ in the usual notation~\cite{fide}.

A typical case for the relevant intensities, indicating clearly the
novel CPTV $\omega$-effects, is depicted in Fig.~\ref{intensomega}.

As announced, the novel $\omega$-effect appears in
the combination $\frac{|\omega|}{|\eta_{+-}|}$, thereby implying
that the decay channel to $\pi^+\pi^-$ is particularly sensitive to
the $\omega$ effect~\cite{bmp}, due to the enhancement by
$1/|\eta_{+-}| \sim 10^{3}$, implying sensitivities up to
$|\omega|\sim 10^{-6}$ in $\phi$ factories. The physical reason for
this enhancement is that $\omega$ enters through $K_SK_S$ as opposed to
$K_LK_S$ terms, and the $K_L \to \pi^+\pi^-$ decay is CP-violating.

\subsection{Microscopic Models for the $\omega$-Effect and Order-of-Magnitude
Estimates}

For future experimental searches for the $\omega$-effect it is
important to estimate its expected order of magnitude, at least  in
some models of foam.

\begin{figure}[htb]
\centering
 \epsfig{file=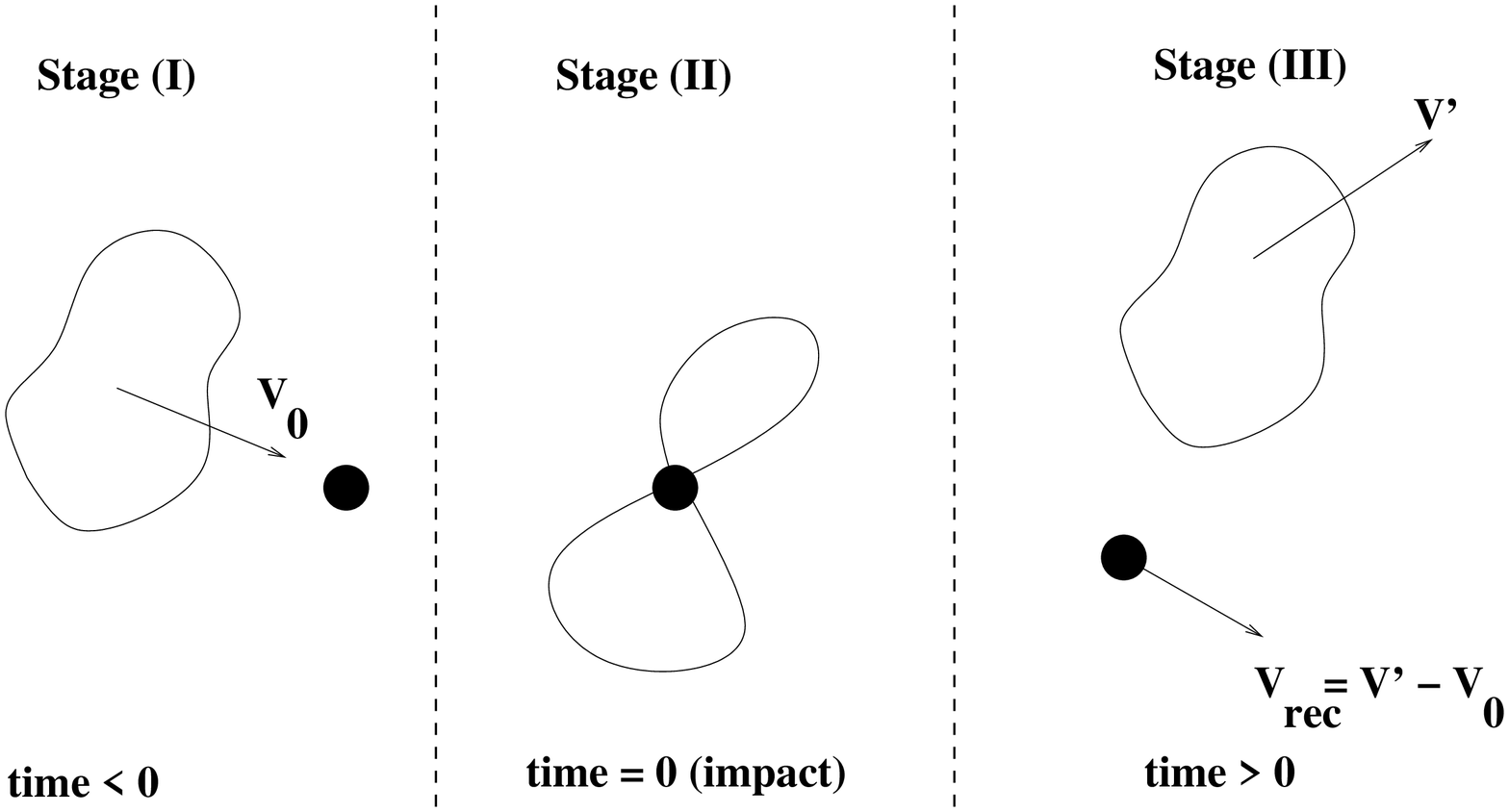, width=0.4\textwidth}
\hfill \epsfig{file=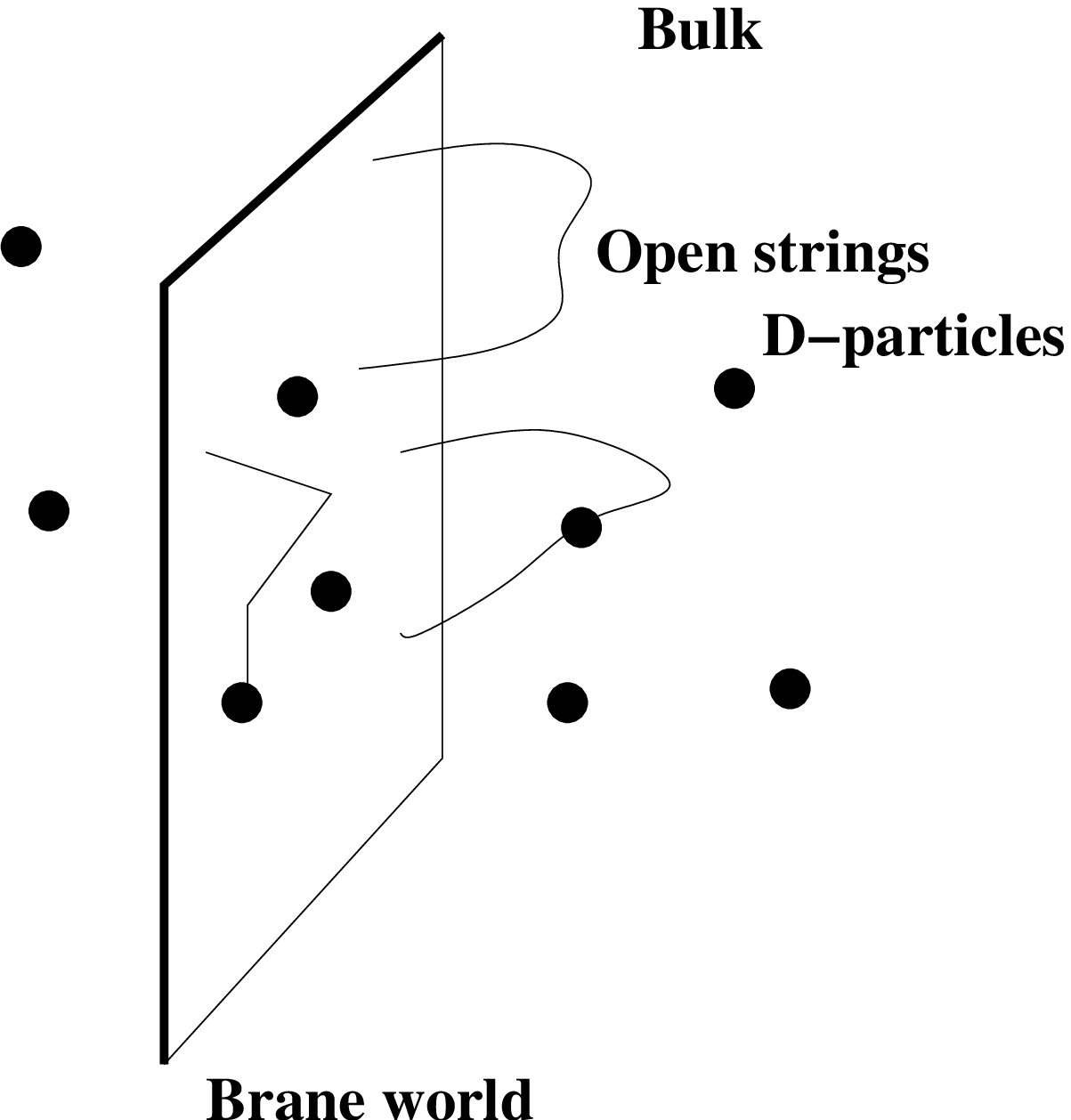, width=0.2\textwidth}
\caption{\underline{Upper}: Recoil of closed string states with
D-particles (space-time defects). \underline{Lower}: A
supersymmetric brane world model of D-particle foam. In both cases
the recoil of (massive) D-particle defect causes distortion of
space-time, stochastic metric fluctuations are possible and the
emergent post-recoil string state may differ by flavour and CP
phases.} \label{drecoil}\end{figure}

A specific model is that of the D-particle foam~\cite{emw,ms,bms},
discussed already in connection with the stochastic
metric-fluctuation approach to decoherence. An important feature for
the appearance of an $\omega$-like effect is that, during each
scattering with a D-particle defect, there is (momentary) capture of
the string state (representing matter) by the defect, and a possible
change in phase and/or flavour for the particle state emerging from
such a capture (see Fig.~\ref{drecoil}).

 The induced metric distortions,
including such flavour changes for the emergent post-recoil matter
state, are:

{\scriptsize
\begin{eqnarray*}
&& g^{00} =\left( -1+r_{4}\right) \mathsf{1}~, \nonumber \\
&& g^{01}=g^{10}=r_{0}\mathsf{1}+ r_{1}\sigma_{1}+ r_{2}\sigma_{2}
+r_{3}\sigma_{3}, \nonumber \\
&&  g^{11} =\left( 1+r_{5}\right) \mathsf{1}\nonumber\end{eqnarray*}}where
the $\sigma_i$ are Pauli matrices.

The target-space metric state is the density matrix $\rho_{\mathrm{grav}}$
defined at (\ref{gravdensity})~\cite{bms}, with the same assumptions for
the parameters $r_{\mu}$ stated there.
The order of magnitude  of the metric elements $g_{0i}\simeq
\overline{v}_{i,rec} \propto g_s\frac{\Delta p_{i}}{M_{s}} $, where
$\Delta p_{i}\sim {\tilde \xi} p_i $ is the momentum transfer during
the scattering of the particle probe (kaon) with the D-particle
defect, $g_{s}<1$ is the string coupling, assumed weak, and $M_{s}$
is the string scale, which in the modern approach to string/brane
theory is not necessarily identified with the four-dimensional
Planck scale, and is left as a phenomenological parameter to be
constrained by experiment.

To estimate the order of magnitude of the $\omega$-effect we
construct the gravitationally-dressed initial entangled state using
stationary perturbation theory for degenerate states~\cite{bmp}, the
degeneracy being provided by the CP-violating effects. As
Hamiltonian function we use

{\scriptsize \begin{eqnarray*}
\widehat{H}=g^{01}\left( g^{00}\right)
^{-1}\widehat{k}-\left( g^{00}\right) ^{-1}\sqrt{\left(
g^{01}\right) ^{2}{k}^{2}-g^{00}\left( g^{11}k^{2}+m^{2}\right)  }
\nonumber
\end{eqnarray*}}describing propagation in the above-described
stochastically-fluctuating space-time. To leading order in the
variables $r$ the interaction Hamiltonian reads:
\begin{equation}
\widehat{H_{I}} = -\left(  { r_{1} \sigma_{1} + r_{2} \sigma_{2}}
\right) \widehat{k} \nonumber
\end{equation}
with the notation {\small $\left| K_{L}\right\rangle =\left|
\uparrow\right\rangle~, \quad \left| K_{S}\right\rangle =\left|
\downarrow\right\rangle .$} The gravitationally-dressed initial states
then can be constructed using stationary perturbation theory:

{\scriptsize
\begin{eqnarray*} \left|  k^{\left(  i\right)
},\downarrow\right\rangle _{QG}^{\left( i\right)  } =  \left| k^{\left(
i\right) },\downarrow\right\rangle ^{\left( i\right)  } + \left| k^{\left(
i\right) },\uparrow\right\rangle ^{\left( i\right) } \alpha^{\left(
i\right)  }~, \nonumber
\end{eqnarray*}}where {\small $ \alpha^{\left(  i\right)  }= \frac{^{\left( i\right)
}\left\langle \uparrow, k^{\left( i\right) }\right| \widehat{H_{I}}\left|
k^{\left(  i\right)  }, \downarrow\right\rangle ^{\left(  i\right)
}}{E_{2} - E_{1}} $}. For $\left|  { k^{\left(  i\right) }, \uparrow}
\right\rangle ^{\left( i \right)  } $  the dressed state is obtained by
$\left| \downarrow\right\rangle \leftrightarrow\left|
\uparrow\right\rangle $ and $\alpha\to\beta$ where  {\small $
\beta^{\left( i\right) }= \frac{^{\left( i\right)  }\left\langle
\downarrow, k^{\left( i\right)  }\right| \widehat{H_{I}}\left| k^{\left(
i\right)  }, \uparrow\right\rangle ^{\left(  i\right) }}{E_{1} - E_{2}}$}.

The totally antisymmetric ``gravitationally-dressed'' state of two mesons
(kaons) is then:

{\scriptsize
\begin{eqnarray*}
\begin{array}
[c]{l}%
\left|  {k, \uparrow} \right\rangle _{QG}^{\left(  1 \right)  }
\left|  { - k, \downarrow} \right\rangle _{QG}^{\left(  2 \right)  }
- \left|  {k, \downarrow} \right\rangle _{QG}^{\left(  1 \right)  }
\left|  { - k, \uparrow} \right\rangle _{QG}^{\left(  2 \right)  }
= \\
\left|  {k, \uparrow} \right\rangle ^{\left(  1 \right) } \left|
{ - k, \downarrow} \right\rangle ^{\left(  2 \right)  } - \left| {k,
\downarrow} \right\rangle ^{\left(  1 \right)  } \left| { - k,
\uparrow} \right\rangle
^{\left(  2 \right)  }\\
+  \left|  {k, \downarrow} \right\rangle ^{\left(  1 \right) }
\left|  { - k, \downarrow} \right\rangle ^{\left(  2 \right)  }
\left(  {\beta^{\left(  1 \right)  } - \beta^{\left(  2 \right)  } }
\right) + \\
\left|  {k, \uparrow} \right\rangle ^{\left( 1 \right)
} \left|  { - k, \uparrow} \right\rangle ^{\left(  2 \right)  }
\left( {\alpha^{\left(  2 \right)  } - \alpha^{\left(
1 \right)  } } \right) \\
 + \beta^{\left(  1 \right)  } \alpha^{\left(  2 \right) }
\left|  {k, \downarrow} \right\rangle ^{\left(  1 \right) } \left| {
- k, \uparrow} \right\rangle ^{\left(  2 \right)  } - \alpha^{\left(
1 \right)  } \beta^{\left( 2 \right)  } \left|  {k, \uparrow}
\right\rangle ^{\left(  1
\right)  } \left|  { - k, \downarrow} \right\rangle ^{\left(  2 \right)  }~.\\
\label{entangl}%
\end{array}\end{eqnarray*}}Notice here
that, for our order-of-magnitude estimates, it suffices to assume that the
initial entangled state of kaons is a pure state. In practice, due to the
omnipresence of foam, this may not be entirely true, but this should not
affect our order-of-magnitude estimates based on such an assumption.

With these remarks in mind we then write for the initial state of
two kaons after the $\phi$ decay:

{\scriptsize  \begin{eqnarray*}
&&\left| \psi\right\rangle = \left|  k,\uparrow\right\rangle
^{\left( 1\right) }\left| -k,\downarrow\right\rangle ^{\left(
2\right) }-\left| k,\downarrow\right\rangle ^{\left(  1\right)
}\left|
-k,\uparrow \right\rangle ^{\left(  2\right)  }+  \nonumber \\
&& \xi \left| k,\uparrow\right\rangle ^{\left( 1\right) }\left|
-k,\uparrow\right\rangle ^{\left(  2\right) }+  \xi^{\prime} \left|
k,\downarrow\right\rangle ^{\left( 1\right) }\left|
-k,\downarrow\right\rangle ^{\left( 2\right)  }~, \nonumber
\end{eqnarray*}}where for $r_{i} \propto\delta_{i1} $ we have
$\xi= \xi^{\prime}$, that is strangeness violation, whilst for
$r_{i} \propto\delta_{i2}$ $\longrightarrow $ $\xi= -\xi^{\prime}$)
(since $\alpha^{\left( i \right) } = \beta^{\left( i \right) } )$ we
obtain a strangeness conserving $\omega$-effect.

Upon averaging the density matrix over $r_{i}$, only the
$|\omega|^{2}$ terms survive: {\small \begin{eqnarray*} &&
|\omega|^{2} = \mathcal{O}\left(  \frac{1}{(E_{1} - E_{2})^2}
(\langle \downarrow, k |H_{I} |k, \uparrow\rangle)^{2} \right)  \sim
\nonumber
\\ && \frac{\Delta_{2} k^{2}}{(m_{1} - m_{2})^{2}} \nonumber
\end{eqnarray*}}for momenta of order of the rest energies, as is the case of a
$\phi$ factory.

Recalling that in the recoil D-particle model under consideration we
have~\cite{emn,bms} $\Delta_{2} = {\tilde \xi}^{2} k^{2}/M_{P}^{2}$,
we obtain the following order of magnitude estimate of the $\omega$
effect: {\small \begin{eqnarray} |\omega|^{2} \sim\frac{{\tilde
\xi}^{2} k^{4}}{M_{P}^{2} (m_{1} - m_{2})^{2}}~. \label{orderomega}
\end{eqnarray}}For neutral kaons with momenta of the order of the rest energies
$|\omega| \sim10^{-4} |{\tilde \xi}|$. For $1 >
{\tilde\xi}\ge10^{-2}$ this not far below the sensitivity of current
facilities, such as KLOE at DA$\Phi$NE. In fact, the KLOE experiment
has just released the first measurement of the $\omega$
parameter~\cite{adidomenico}: {\small \begin{eqnarray*} &&
{\rm Re}(\omega) = \left(
1.1^{+8.7}_{-5.3} \pm 0.9\right)\times 10^{-4}~, \nonumber \\
&& {\rm Im}(\omega) = \left( 3.4^{+4.8}_{-5.0} \pm 0.6\right)\times
10^{-4}~. \nonumber\end{eqnarray*}}One can constrain the $\omega$
parameter (or, in the context of the above specific model, the
momentum-transfer parameter ${\tilde \xi}$) significantly in
upgraded facilities. For instance, there are the following
perspectives for KLOE-2 at (the upgraded)
DA$\Phi$NE-2~\cite{adidomenico}: ${\rm Re}(\omega),~ {\rm
Im}(\omega) \longrightarrow 2 \times 10^{-5}$.

Let us now mention that $\omega$-like effects can also be generated by
the Hamiltonian evolution of the system as a result of gravitational
medium interactions. To this end, let us consider the Hamiltonian
evolution in our stochastically-fluctuating D-particle-recoil
distorted space-times, {\scriptsize  $ \left| \psi\left(  t\right)
\right\rangle =\exp\left[ -i\left( \widehat
{H}^{(1)}+\widehat{H}^{\left( 2\right) }\right)
\frac{t}{\hbar}\right] \left| \psi\right\rangle$}.

Assuming for simplicity $\xi = \xi^\prime = 0$, it is easy to
see~\cite{bms} that the time-evolved state of two kaons contains
strangeness-conserving $\omega$-terms:

{\scriptsize
\begin{eqnarray*} && \left|  \psi\left(  t\right)  \right\rangle \sim
e^{-i\left(  \lambda_{0}^{\left(  1\right)  }+
\lambda_{0}^{\left(  2\right)  }\right)  t}%
 \varpi\left(  t\right) \times \nonumber \\
&& \left\{  \left| k,\uparrow\right\rangle ^{\left( 1\right) }\left|
-k,\uparrow\right\rangle ^{\left( 2\right)  }-\left|
k,\downarrow\right\rangle ^{\left(  1\right) }\left|  -k,\downarrow
\right\rangle ^{\left(  2\right)  }\right\}~. \nonumber
\end{eqnarray*}}
The quantity $\varpi (t)$ obtained within this specific model is purely imaginary,

{\scriptsize \begin{eqnarray*} && {\cal O}\left(\varpi\right)  \simeq
i\frac{2\Delta_{1}^{\frac{1}{2}}k}{\left(
k^{2}+m_{1}^{2}\right)^{\frac{1}{2}}-\left(
k^{2}+m_{2}^{2}\right)^{\frac{1}{2}}}\times \nonumber \\
&&\cos\left( \left| \lambda^{\left(  1\right) }\right| t\right)
\sin\left( \left| \lambda^{\left(  1\right) }\right| t\right) =
\varpi_{0}\sin\left( 2\left|  \lambda^{\left( 1\right) }\right|
t\right), \nonumber
\end{eqnarray*}}with $\Delta_{1}^{1/2}\sim\left| {\tilde \xi}\right| \frac{\left| k\right|
}{M_{P}}$,
$\varpi_{0}\equiv\frac{\Delta_{1}^{\frac{1}{2}}%
k}{\left(  k^{2}+m_{1}^{2}\right)  ^{\frac{1}{2}}-\left(  k^{2}+m_{2}%
^{2}\right)  ^{\frac{1}{2}}}$, $\left|  \lambda^{\left(  1\right)
}\right|  \sim\left( 1+\Delta_{4}^{\frac{1}{2}}\right)
\sqrt{\chi_{1}^{2}+\chi_{3}^{2}}$,~ $\chi_{3}\sim\left(
k^{2}+m_{1}^{2}\right)  ^{\frac{1}{2}}-\left(
k^{2}+m_{2}^{2}\right) ^{\frac{1}{2}}$.

It is important to notice the time dependence of the medium-generated
effect. It is also interesting to observe that, if in the initial state
we have  a strangeness-conserving (-violating) combination,
$\xi = -\xi^\prime $  ($\xi = \xi^\prime $),
then the time
evolution generates time-dependent strangeness-violating
(-conserving $\omega$-) imaginary effects.

The above description of medium effects using Hamiltonian evolution is
approximate, but suffices for the purposes of obtaining order-of-magnitude
estimates for the relevant parameters. In the complete description of the
above model  there is of course decoherence~\cite{bms,emn}, which affects
the evolution and induces mixed states for kaons. A complete analysis of
both effects, $\omega$-like and decoherence in entangled neutral kaons of
a $\phi$-factory, has already been carried out~\cite{bmp}, with the upshot
that the various effects can be disentangled experimentally, at least in
principle (see Section 4.6 below).

Finally, as the analysis of
\cite{bms} demonstrates, no $\omega$-like effects are generated by
thermal bath-like (rotationally-invariant, isotropic) space-time
foam situations, argued to simulate the QG
environment in some models \cite{garay}. In this
way, the potential observation of an $\omega$-like effect in
EPR-correlated meson states would in principle distinguish various
types of space-time foam.

\subsection{Disentangling the $\omega$-Effect from
the C(even)  Background}

When interpretating experimental results on delicate violations of CPT
symmetry, it is important
to disentangle (possible) genuine effects from those
due to ordinary physics.
Such a situation arises in connection with the $\omega$-effect,
that must be disentangled from the C(even) background
characterizing the decay products in a
$\phi$-factory~\cite{dunietz}.

The C(even) background $e^+e^- \Rightarrow 2\gamma \Rightarrow K^0
{\overline K}^0$ leads to states of the form

{\scriptsize\begin{eqnarray*}
 &&|b> = |K^0 {\overline K}^0 (C({\rm even}))> =\nonumber\\
&& \frac{1}{\sqrt{2}}\left(K^0({\vec k}) {\overline K}^0 (-{\vec k})
+  {\overline K}^0({\vec k}) K^0 (-{\vec k}) \right)~, \nonumber
\end{eqnarray*}}which at first sight mimic the $\omega$-effect, as such states would
also produce contamination by terms $K_SK_S,~K_LK_L$.

Closer inspection reveals, however,
that the two types of effects can be clearly disentangled
experimentally. The reason is two-fold.

(i) First of all, the order of magnitude of the C(even) background
is \emph{much smaller} than the C(odd) resonant contribution, as we
have seen in  the previous discussion, at least in the context of a
class of models~\cite{bms}. Indeed, unitarity
bounds~\cite{dunietz,dafne2} imply for the C(even) background:
{\small \begin{eqnarray*} \frac{\sigma (e^+e^- \to K^0{\overline
K}^0, J^P=0^+)}{\sigma(e^+e^- \to \phi \to K_SK_L ) } \ge 3.6 \times
10^{-10}~, \nonumber \end{eqnarray*}}and actually one expects the
inequality to be saturated. In contrast, the order of magnitude of
the $\omega$-effect might be much larger, at least in some models
(\ref{orderomega}).

(ii) A more important feature, which clearly distinguishes the
$\omega$-effect from the ``fake'' background effects, is its
\emph{different interference} with the C(odd) background~\cite{bmp}.
Terms of the type $K_S K_S$ (which dominate over $K_L K_L$) coming
from the $\phi$-resonance as a result of $\omega$-CPTV can be
distinguished from those coming from the $C=+$ (even) background
because they interfere differently with the regular $C=-$ (odd)
resonant contribution with $\omega=0$.

\begin{figure}
\centering
  \epsfig{file=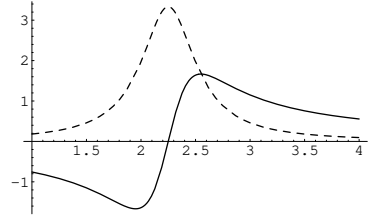, width=0.3\textwidth}
\caption{Disentangling the $\omega$-effect from C(even) background:
different behaviour at the resonance. The $C=-$ contribution (solid
line) vanishes at the top of the resonance, while the C=+ genuine
effect (dashed line) still exhibits a resonance peak.}
\label{resonance} \end{figure}

Indeed, in the CPTV case, the $K_L K_S$ and  $\omega K_S K_S$ terms
have the same dependence on the center-of-mass energy $s$ of the
colliding particles producing the resonance, because both terms
originate from the  $\phi$-particle. Their interference, therefore,
being proportional to the real part of the product of the
corresponding amplitudes, still displays a peak at the resonance.

On the other hand, the amplitude of the $K_S K_S$ coming from the
$C=+$ background has no appreciable dependence on $s$ and has
practically vanishing imaginary part. Therefore, given that the real
part of a Breit-Wigner amplitude vanishes at the top of the
resonance, this implies that the interference of the $C=+$
background with the regular $C=-$ resonant contribution vanishes at
the top of the resonance, with opposite signs on both sides of the
latter (see Fig~\ref{resonance}). This clearly distinguishes
experimentally the two cases.

\subsection{Disentangling the $\omega$-Effect from Decoherent
Evolution Effects}

As a final point in this section we discuss briefly the experimental
disentanglement of the $\omega$-effect from decoherent evolution
effects~\cite{bmp}.

In models of space-time foam, the initial entangled state of two
kaons, after the $\phi$-meson decay, is actually itself a density
matrix  $\tilde{\rho}_{\phi} ={\rm Tr}|\phi><\phi|$. For $\omega =
0$, the density matrix assumes the form (we remind the reader that
the requirement of complete positivity in the entangled-kaon case
implies~\cite{benatti} that the decoherent coefficients are $\alpha
= \gamma, \beta = 0$)~\cite{huet}:

{\scriptsize
\begin{eqnarray*} &&\tilde{\rho}_{\phi} = \rho_{S} \otimes \rho_{L} +
\rho_{L} \otimes \rho_{S} - \rho_{I}\otimes \rho_{{\overline I}} -
\rho_{{\overline I}}\otimes \rho_{I}
\nonumber\\
&& - \frac{i\alpha}{\Delta m} ( \rho_{I}\otimes \rho_I -
\rho_{{\overline I}} \otimes  \rho_{{\overline I}})
-\frac{2\gamma}{\Delta \Gamma} (\rho_S \otimes \rho_S - \rho_L
\otimes \rho_L)~, \nonumber
\end{eqnarray*}}where $\rho_{S} = |S><S|, ~\rho_L = |L><L|, ~\rho_I = |S><L|,
~\rho_{{\overline I}} = |L><S|$, and an overall multiplicative
factor of $\frac{1}{2} \frac{(1 + 2|\epsilon|^2)} {1 -
2|\epsilon|^2{\rm cos}(2\phi_\epsilon)}$ has been suppressed.

Now, for $\omega \ne 0$ but $\gamma = 0$ the initial entangled state
becomes~\cite{bmp}:

{\scriptsize \begin{eqnarray*} \rho_\phi &=& \rho_S \otimes \rho_L +
\rho_L \otimes \rho_S - \rho_{I}\otimes \rho_{{\overline I}} -
\rho_{{\overline I}}\otimes \rho_I
\nonumber\\
&-& \omega (\rho_I \otimes \rho_S - \rho_S \otimes \rho_I ) -
\omega^* (\rho_{{\overline I}} \otimes \rho _S - \rho_S \otimes
\rho_{{\overline I}} )
\nonumber\\
&-& \omega ( \rho_{{\overline I}} \otimes \rho_L - \rho_L \otimes
\rho_{{\overline I}}) - \omega^* ( \rho_I \otimes \rho_L - \rho_L
\otimes \rho_I )
\nonumber\\
&-& |\omega|^2 ( \rho_{I}\otimes \rho_I + \rho_{{\overline I}}
\otimes  \rho_{{\overline I}}) + |\omega|^2 (\rho_S \otimes \rho_S +
\rho_L \otimes \rho_L) \nonumber
\end{eqnarray*}}with the same suppressed multiplicative factor as in the
previous equation.

The experimental disentanglement of $\omega$ from the decoherence
parameter $\gamma$ is possible as a result of different symmetry
properties and different structures generated by the time evolution
of the pertinent terms. A detailed phenomenological analysis in
various channels for $\phi$ factories has been performed in
\cite{bmp}, where we refer the interested reader for details.

\section{Precision T, CP and CPT Tests with Charged Kaons}

It turns out that
precision tests of discrete symmetries can also be performed with charged
kaons. This realization
has generated great interest~\cite{ali},  mainly due to
the (recently acquired) high statistics of the NA48
experiment~\cite{NA48} in certain decay channels. In fact,
as we will argue in this section,
while the primary objective of this experiment is to
probe in detail certain
aspects of chiral perturbation theory,
it could also furnish strong constraints for various
new physics scenarios.

For the purpose of testing CPT symmetry we shall restrict ourselves
to one particular charged kaon decay, $K^\pm \rightarrow \pi^+ +
\pi^- + \ell ^\pm + \nu_\ell (\overline \nu_\ell)$, abbreviated as
$K_{\ell 4}^\pm$.
The CPT symmetry can be tested with this
reaction~\cite{wu,treiman} by comparing the decay rates of $K^+$
with the corresponding decays of the $K^-$ mode.

If CPT is well-defined but does not commute with the Hamiltonian, we
have the relations: $|K^+\rangle =\widehat{CPT}|K^-\rangle $, $
|\pi^+\rangle =\widehat{CPT}|\pi^-\rangle, \quad |\pi^0\rangle =
\widehat{CPT}|\pi^0\rangle $. If CPT does not commute with
the Hamiltonian, then differences between particle antiparticle masses
may occur, but this is not the end of the story. In fact, as
emphasized earlier, this is not true in certain models of Lorentz-
and/or unitarity-violating QG~\cite{ehns,lopez,huet,sme}.

If, on the other hand, CPT is ill-defined, as is the case of
QG-induced  decoherence, then there are (perturbative) ambiguities
in the antiparticle state, which is still well-defined but with
modified properties (see previous section).
However, in contrast to the neutral kaon case, the two charged pions
in this decay are already distinguishable by means of their
electromagnetic interactions (charge), which are, of course,
much stronger than
their (quantum) gravitational counterparts. Hence, in this
respect the ill-defined nature of the CPT operator is not relevant.

A breaking of CPT through unitarity violations (e.g., non-hermitean
effective Hamiltonians) could lead in principle to different decay
widths for the two decay modes $K^\pm$. This would constitute a
straightforward precision test of CPT symmetry, if sufficiently high
statistics for charged kaons were available~\cite{wu,treiman}.
Unlike the entangled neutral kaon case, however,
such
tests could not distinguish between the various types of CPT
breaking.

We next proceed to review briefly the precision tests of T, CP and
CPT symmetry using $K_{\ell 4}^\pm$ decays. With the exception of
tests of T-odd triple correlations that we present at the end of this
section, the discussion parallels that of \cite{wu,treiman}, where we
refer the reader for further details.

It is important to stress once more that in QG, especially in
stochastic space-time foam models, the CPTV is essentially
a microscopic Time Reversal (T) Violation, independent of CP
properties. It is therefore important to discuss precision tests of
T symmetry independent of CP, CPT.

By using $K_{\ell 4}^\pm$ for precision tests of
T, CP, CPT, one can check in parallel the validity of the
$\Delta S = \Delta Q$ rule, exploiting the high statistics a
of the NA48 experiment~\cite{NA48}, e.g., one can look for
the $\Delta S = -\Delta Q$ reaction: $K^+ \to \pi^+ + \pi^+ + e +
\nu $, whose non-observation would establish stringent bounds on the
violation of the rule.

In our analysis below, following \cite{wu,treiman} we assume
the $|\Delta I|=1/2$ isospin rule, which, by the way, can be checked
experimentally, as we shall see.

We use the following notation for the corresponding amplitudes:

{\scriptsize \begin{eqnarray*} && e^{i\xi}A = \nonumber \\ &&
\langle \pi^+\pi^-|\ell=0, m=0 \rangle \langle \ell=0, m=0|S_z +
ivS_4|K^+
\rangle \times \nonumber \\ && m_K^{5/2}(\omega_+\omega_-)^{1/2}~, \nonumber \\
&& e^{i\xi + i\eta_0}B_0 {\rm cos}\theta  = \nonumber \\ && \langle
\pi^+\pi^-|\ell=1, m=0 \rangle \langle \ell=1, m=0|S_z + ivS_4|K^+
\rangle \times \nonumber \\ && m_K^{5/2}(\omega_+\omega_-)^{1/2}~,
\nonumber \\
&& e^{i\xi + i\eta_\pm}B_\pm {\rm sin}\theta e^{\pm i\phi} =
\nonumber \\ && \langle \pi^+\pi^-|\ell=1, m=\pm 1 \rangle \langle
\ell=1, m=\pm
1|\frac{1}{\sqrt{2}}(S_x + iS_y)|K^+ \rangle \times \nonumber \\
&& m_K^{5/2}(\omega_+\omega_-)^{1/2}~, \nonumber
\end{eqnarray*}}with $\ell$ the orbital
angular momentum quantum number, $m$ its
z-axis component. The phase conventions are chosen such that
$A,B_0, B_\pm $ are real and positive; the polar angles
$\theta,\phi$ pertain to the di-pion center-of-mass system
$\Sigma_{2\pi}$ and $x,y,z$ are Cartesian coordinates in the
laboratory system $\Sigma_{\rm Lab}$. An angle
$\alpha$ in the di-lepton center-of-mass system $\Sigma_{\rm \ell
\nu_\ell}$ will also be used. The $\omega_\pm$ denote the
laboratory energies of the $\pi^\pm$, $v=-[m_K - (\vec P^2 +
M^2)^{1/2}]^{-1}|\vec P|$ is the velocity of Lorentz transformation
connecting $\Sigma_{\rm \ell \nu_\ell}$ to $\Sigma_{\rm Lab}$
frames, with $\vec P$ the total momentum of the two pions in
$\Sigma_{\rm Lab}$.

The action of CPT is obtained by replacing the corresponding
amplitudes by barred quantities: $\overline {(...)}$, and implementing the
following substitutions: $K^+ \to K^-,\quad \pi^+(\vec k_1) \pi^-(\vec
k_2) \to \pi^-(\vec k_1) \pi^+(\vec k_2)$, plus appropriate complex
conjugates.

We outline below various possible precision tests of discrete symmetries
based on the reaction $K_{\ell 4}^\pm$:

\begin{itemize}
\item{} \underline{CPT invariance implies}:

$A=\overline A, \quad B_0=\overline B_0, \quad B_\pm = \overline
B_\mp$, \qquad $\eta_+ + \overline \eta_- = \eta_- + \overline
\eta_+ = \eta_0 + \overline \eta_0 = 2\left(\delta_p -
\delta_s\right)$, \underline{independently} of T invariance, with
$\delta_p (\delta_s)$ the strong-interaction $\pi-\pi$ scattering
phase shifts for the states $I=1, \ell = 1 (I=0,\ell = 0)$.

Also, CPT invariance, independently of the $|\Delta I|=1/2$ rule,
implies: $ {\rm rate}(K^+ \to \pi^+\pi^-e^+\nu_e) + {\rm rate}(K^+
\to \pi^0\pi^0e^+\nu_e) = {\rm rate}(K^- \to \pi^+\pi^-e^-\overline
\nu_e) + {\rm rate}(K^- \to \pi^0\pi^0e^-\overline \nu_e)~.$

Under the assumption of the $|\Delta I|=1/2$ rule, on the other
hand, CPT invariance implies for the differential rates $d^5N$ of
$K_{\ell 4}^\pm$:

$\int d\phi d{\rm cos}\theta \quad d^5N(K^+ \to \pi^+\pi^-e^+\nu_e)
= \int d\phi d{\rm cos}\theta \quad d^5N(K^-\to
\pi^+\pi^-e^-\overline \nu_e)~.$

\item{} \underline{T invariance } (independent of CP, CPT) implies:

$\eta_\sigma = \delta_p - \delta_s ~({\rm modulo}\quad \pi)~, \qquad
\sigma = 0, \pm ~.$

\item{} \underline{CP Invariance} (independent of T, CPT), which in
terms of the angles means: $\theta,\phi,\alpha \to \theta, -\phi,
\alpha$, implies:

$A = \overline A, \quad B_0 = \overline B_0, \quad B_\pm = \overline
B_\mp, \quad \eta_0 = \overline \eta_0, \quad \eta_\pm = \overline
\eta_\mp$, and for the differential rates
$[d^5N(K^+)]_{\alpha,\theta,\phi}
=[d^5N(K^-)]_{\alpha,\theta,-\phi}$, leading also to $\int d\phi
d{\rm cos}\theta \quad d^5N(K^+ \to \pi^+\pi^-e^+\nu_e) = \int d\phi
d{\rm cos}\theta \quad d^5N(K^-\to \pi^+\pi^-e^-\overline \nu_e)$
but \emph{without} the assumption about $|\Delta I|=1/2$.

\end{itemize}

Let us now deviate sligtly from the main scope of this article,
and comment on the possibility of testing physics beyond the Standard Model (SM)
 by looking for T-odd triple correlators~\cite{triple} in
the NA48 data for the reaction modes $K_{\ell 4}^\pm$.
Such tests
are not directly related to CPT  but rather
to different aspects of new physics, such as supersymmetry;
the latter, in turn, could be
essential for formulating  consistent theories of QG.

Within the SM, direct CP violation or CP violation of pure
$\Delta S = 1$ origin, which, due to CPT symmetry, would imply
T-odd correlators~\footnote{It should be stressed at this point
that, on account of the anti-unitarity of the time-reversal
operator, $T$-odd correlators are not necessarily $T$-violating.},
is very strongly suppressed in non-leptonic decays $K^\pm \to
(3\pi)^\pm $: ${\cal O}\left(10^{-5}-10^{-6}\right)$ and absolutely
negligible in $K_{\ell 4}$~\cite{AmbrosioIsidori}. Evidence for such
violations in $K_{\ell 4}$ charged-kaon decays would, therefore,
constitute evidence for physics beyond the SM.

As was pointed out in~\cite{triple}, one can construct
appropriate CP observables for charged kaon decays $K_{\ell 4}$
that do not involve the lepton polarization, a quantity difficult
to measure in the NA48 experiment~\cite{NA48}. This is
achieved by considering appropriate combinations of  matrix elements
pertaining to \emph{both} decay modes $K_{\ell 4}^\pm$. The
construction makes use of the fact that, under the assumption of
only left-handed neutrinos, the most general local effective
Hamiltonian, relevant to charged-kaon $K_{\ell
4}$ (and $K_{\ell 3}$) decays, can be expanded in terms of
appropriate local dimension six field operators ${\cal
O}_i$~\cite{triple}: {\small
\begin{eqnarray*} {\cal H}_{\rm eff} = 2\sqrt{2}G_F V_{us}^*\sum_i
C_i {\cal O}_i + h.c.~,
\end{eqnarray*}}where the operators ${\cal O}_i$ are four-fermion operators
involving left-handed neutrinos (e.g. ${\cal O}^V_L = \overline
s_L\gamma^\mu u_L \overline \nu_L \gamma_\mu \ell_L,~ {\cal O}^S_L =
\overline s_R u_L \overline \nu_L \ell_R,~ {\cal O}^T_L = \overline
s_R \sigma^{\mu\nu}u_L \overline \nu_L \sigma_{\mu\nu} \ell_R$, etc
(${\cal O}^i_R: s_R \to s_L,~u_L \to u_R$)). In the SM only
$C_L^V=1$, while the others are negligible.

Within the SM, the relevant matrix elements for the $K_{\ell 4}$
decay are

{\small \begin{eqnarray*} \langle \pi^+ \pi^- |\overline s
\gamma^\mu u |K^+\rangle , \langle \pi^+ \pi^- |\overline s
\gamma^\mu \gamma_5 u |K^+\rangle~. \end{eqnarray*}}Beyond the SM,
one has more structures; for instance~\cite{triple} {\small
\begin{eqnarray*} \langle \pi^+ \pi^- |\overline s \gamma_5 u
|K^+\rangle , \langle \pi^+ \pi^- |\overline s \sigma^{\mu\nu}
\gamma_5 u |K^+\rangle~.\end{eqnarray*}}Using such structures, one
can construct~\cite{triple} appropriate combinations of T-odd
correlators in $K_{\ell 4}$ decays, proportional to momentum triple
products, $\vec p_\ell \cdot (\vec p_{\pi_1} \times \vec
p_{\pi_2})$, by using \emph{both} $K^+$ and $K^-$ modes. This leads
to new CP-violating observables, free from strong final-state
interactions. These observables can be used for precision CP tests
without the need of measuring lepton polarization. The
results are complementary to those
obtained through normal-to-the-decay-plane
muon polarization in $K_{\mu 3}$ decays, and of comparable accuracy.
For details and related references we refer the interested reader to
the literature~\cite{triple}.

We close this section by pointing out that
the NA48 data could
also provide new
stringent constraints on exotic (beyond the SM)
scenarios, such as R-parity breaking in supersymmetric theories,
complementary to those obtained through $K_{\ell 3}$ or other
decays. In fact, as has been known for some time~\cite{christova},
the existence of complex coupling constants
allows to test supersymmetry in weak decays (in particular rare kaon
decays involving leptons). Specifically, T-invariance may be studied
with appropriate T-odd observables, such as triple
correlators of polarizations and momenta (for instance, in $K_{\mu
3}^+$ decays the appropriate observable is the normal-to-the-decay-plane
muon
polarization $\langle \vec \sigma_\mu \cdot (\vec p_\mu \times \vec
p_\pi)/|\vec p_\mu \times \vec p_\pi|\rangle$). This type of analyses can be
complemented by the above-mentioned study of
lepton-polarization-independent T-odd observables in $K_{\ell
4}^\pm$ decays, and also serve as precision tests of CPT symmetry.
To the best of our knowledge this has not been done yet.

\section{Instead of Conclusions}

In this review we have outlined several aspects of CPTV and
the corresponding experiments. We have attempted to convey
a general feel for the interesting and challenging precision tests
that can be performed using kaon systems. Such experiments could shed light
on many aspects of an extended class of QG
models, featuring decoherence of low-energy matter
due to its propagation in foamy backgrounds.

We hope to have presented sufficient theoretical motivation and
estimates of the associated effects to support the case that
testing QG experimentally at present
fascilities may turn out to be a worthwhile endeavour.
In fact, as we have argued, CPTV may be a real feature of QG,
that can be tested and observed, if true, in the foreseeable future.

We have outlined various,
schemes for CPT breaking, that are in
principle independent.
We have stressed that decoherence and Lorentz
violation (LV) are independent effects: in QG one may have
Lorentz-invariant  (LI) decoherence~\cite{Millburn}.
The frame dependence of LV effects (e.g.
day-night differences) could serve to disentangle LV from LI CPTV.
The example discussed in this article is a comparison between results
in meson factories. If there is LV, then there should in principle
be frame-dependent differences between $\phi$-factories, where
the initial meson is produced at rest, and $B$-meson factories, where
the initial $\Upsilon$-state is boosted.

As mentioned above, precision tests of fundamental symmetry in meson
factories could provide sensitive probes of QG-induced decoherence
and CPTV. In particular, one might observe novel effects ($\omega$-effects)
\underline{exclusive} to entangled neutral meson states, modified
EPR correlations, and, as a consequence, theoretical (intrinsic)
limitations on flavour tagging for $B$-factories~\cite{bomega}. As
we have seen, some theoretical (string-inspired) models of
space-time foam predict $\omega$-like effects of an order of
magnitude that is already well within the reach of the next upgrade of
$\phi$-factories, such as DA$\Phi$NE-2.

Precision experiments to test discrete space-time symmetries can
also be performed with charged kaons: the pertinent
experiments~\cite{NA48} can carry out high-precision tests of T, CPT
and CP invariance, including aspects of physics beyond the Standard
Model, such as supersymmetry, using $K_{\ell 4}$ decays.

The current experimental situation for QG
signals appears exciting,
and several experiments are reaching interesting regimes, where many
theoretical models can be falsified. More precision experiments are
becoming available, and many others are being designed for the
immediate future. Searching for tiny effects of this elusive theory
may at the end be very rewarding. Surprises may be around the
corner, so it is worth investing time and effort. Nevertheless, much
more work, both theoretical and experimental, needs to be done
before (even tentative) conclusions regarding QG
effects are reached.

\section*{Acknowledgements}

We thank A. Di Domenico for the invitation to write this review and for
many illuminating discussions. We also thank E. Alvarez, G.
Amelino-Camelia, F. Botella, M. Nebot, Sarben Sarkar, A.~Waldron-Lauda,
and M.~Westmuckett  for discussions and collaboration on some of the
topics reviewed here; J.B. and N.E.M. thank B. Bloch-Devaux, L. Di Lella,
G. Isidori and B. Peyaud for informative discussions on charged-kaon
decays. The work of J.B. and J.P. is supported by  Spanish MEC and
European FEDER under grant FPA 2005-01678. The work of D.V.N. is supported
by D.O.E. grant DE-FG03-95-ER-40917.

\end{document}